\documentclass[twocolumn,showpacs,prl,aps,nofootinbib,showkeys,unsortedaddress,10pt]{revtex4-2}

\usepackage{amsmath}
\usepackage{amssymb}
\usepackage{slashed}
\usepackage{hyperref}
\usepackage{color}
\usepackage{comment}
\usepackage{simpler-wick}
\usepackage{graphicx}	
\usepackage[english]{babel}
\usepackage{amsmath,amsfonts,amsthm,amssymb, psfrag} 
\usepackage{yfonts}
\usepackage[fleqn,tbtags]{mathtools}
\usepackage{tikz}
\usepackage{tikz-cd}
\usetikzlibrary{decorations.markings}
\usepackage{hyperref}
\usepackage[capitalise]{cleveref}
\usepackage{pgfplots}
\usepackage{colortbl}
\usepackage{physics}

\def\p{\partial}

\usepackage{wrapfig}

\usepackage{simplewick}
\usepackage{slashed}
\def\be{\begin{equation}}
\def\ee{\end{equation}}
\def\bea{\begin{eqnarray}}
\def\eea{\end{eqnarray}}

\newcommand{\oll}{\overline}

\newcommand{\Cset}{{\,\,{{{^{_{\pmb{\mid}}}}\kern-.47em{\mathrm C}}}}}

\newcommand{\half}{\frac12}

\allowdisplaybreaks

\usepackage{tikz-feynhand}

\newcommand{\psib}{\Bar{\psi}}

\definecolor{darkgreen}{rgb}{0,.7,0}
\definecolor{linkblue}{rgb}{0.,0.,0.9333}

\hypersetup{
    pdffitwindow=true,      
    pdfauthor={J. Bath},     
    pdfnewwindow=true,      
    bookmarksnumbered=true,
    colorlinks=true,      
    linkcolor=red,          
    citecolor=darkgreen,        
    filecolor=magenta,      
    urlcolor=cyan,           
    menucolor=blue,
}

\AtBeginDocument{\setlength\abovedisplayskip{5pt}}
\AtBeginDocument{\setlength\belowdisplayskip{5pt}}

\begin{document}

\title{Systematic Analytic Regularization in $\varphi^4$ and Yukawa Theories}

\author{J. Bath}
\email{jarryd.bath@tuks.co.za}
\affiliation{
 Department of Physics, University of Pretoria,
Private Bag X20, Hatfield 0028, South Africa
}
\author{W. A. Horowitz}
\email{wa.horowitz@uct.ac.za}
\affiliation{Department of Physics, University of Cape Town, Rondebosch 7701, South Africa, Department of Physics, New Mexico State University, Las Cruces, New Mexico, 88003, USA  and\\
Theoretical Sciences Visiting Program,
Okinawa Institute of Science and Technology Graduate University, Onna, 904-0495, Japan}
\date{\today}

\begin{abstract}
    We introduce a novel regularization scheme: Systematic Analytic Regularization (SAR). SAR regularizes a theory at the level of the action by analytically continuing the power of the kinetic operator, ensuring that the theory is formally finite before any terms in the Dyson series are evaluated. We demonstrate that SAR fully and self-consistently regularizes $\varphi^4$ and Yukawa theories at NLO. 
\end{abstract}
\maketitle

\section{Introduction}
Quantum field theories (QFTs) are a natural framework that results from combining the postulates of special relativity and quantum mechanics and are central to our understanding of fundamental physics \cite{Peskin:1995ev,Sterman:1993hfp,Weinberg:1995mt}. Most notably, the Standard Model of particle physics is a QFT that describes three of the four fundamental forces in nature, and is the most successful description of Nature known \cite{tHooft2007,Peskin:1995ev,Sterman:1993hfp,Weinberg:1995mt,Weinberg:1996kr}. When computing various quantities in quantum field theories, by working in a perturbative expansion of the time ordered exponential, one often encounters integrals that are formally divergent \cite{Peskin:1995ev, Sterman:1993hfp,Weinberg:1995mt,Weinberg:1996kr}. These divergences must be tamed by a regularization procedure, where the integral gains an extra dependence in a variable, known as a regulator \cite{Sterman:1993hfp,Weinberg:1995mt,Weinberg:1996kr}. Renormalization is a procedure that is then applied to cancel these infinities and to connect the parameters in the Lagrangian with experimental measurements \cite{Peskin:1995ev, Sterman:1993hfp,Weinberg:1995mt,Weinberg:1996kr}. After renormalization, the regulator can be smoothly removed with the computed observables remaining finite \cite{Peskin:1995ev, Sterman:1993hfp,Weinberg:1995mt,Weinberg:1996kr}.
\par Several regularization schemes exist, including: momentum cut off \cite{Heisenberg1936}, Pauli-Villars \cite{Pauli:1949zm}, dimensional regularization (dim reg) \cite{tHooft:1979rtg} and supersymmetric dimensional reduction \cite{Siegel:1979wq}, zeta \cite{Hawking:1976ja} and operator regularization \cite{McKeon:1986rc}, analytic regularization \cite{Bollini1964,speer,Lee:1983gj}, and denominator regularization  \cite{Horowitz:2022uak}. Each of these procedures has practical and conceptual advantages and disadvantages. In this work, we introduce a novel regularization scheme that we name systematic analytic regularization (SAR), which builds on analytic regularization. SAR applies analytic continuation directly at the level of the action using tools from fractional calculus \cite{Bollini1964,speer,Lee:1983gj}, aiming to address the shortcomings of existing methods in a systematic and symmetry-respecting manner such that the theory is always finite and avoids the use of \emph{ad hoc} methods.   
\par Let us now discuss some of the advantages and disadvantages of the various regularization schemes we mentioned. Ultraviolet (UV) divergences in QFTs originate from integrals over momenta on an unbounded domain in loop diagrams. The most natural approach to rendering these integrals finite is to impose a momentum cut-off. Such a cut-off could be indicative that the QFT only describes physics up to a certain momentum scale. In this regularization scheme, there is a natural interpretation of the renormalization group (RG) through the interpretation of integrating out higher energy modes \cite{Wilson:1973jj,Peskin:1995ev}. There are two main issues related to imposing a momentum cut-off. First, a momentum cut-off is in direct contradiction to Lorentz invariance. Second, the momentum cut-off scheme violates the Ward identity in quantum electrodynamics (QED) \cite{Peskin:1995ev}.
\par In Pauli-Villars regularization, massive fictitious particles are introduced with statistics such that their associated propagators cancel the divergences that arise from the propagators associated with the physical particles. This procedure naturally preserves Lorentz invariance. However, Pauli-Villars regularization can become impractical when ensuring gauge invariance, as it can necessitate the introduction of an infinite number of fictitious particles \cite{Bjorken:1965zz,Slavnov:1971aw}.
\par Dimensional regularization (dim reg) is the most common regularization procedure used in QFT \cite{tHooft:1973mfk}. In dim reg, the dimensions of the spacetime are analytically continued from an integral number $d$ to $d-\epsilon$, where $d$ is usually 4. By reducing the number of dimensions by $\epsilon$, one lowers the power of the numerator of the integrand in divergent integrals after the usual procedure of switching to spherical coordinates, improving the convergence properties. Dim reg has many advantages: it is relatively easy to implement; gauge invariance is preserved at all orders in perturbation theory \cite{tHooft:1972tcz}; and it aligns naturally with the modified minimal subtraction renormalization scheme ($\overline{\text{MS}}$) \cite{Weinberg:1973xwm}. Furthermore, the introduction of a fictitious scale 
$\mu$ during the continuation allows physical observables to remain dimensionally consistent, enabling the application of the Callan–Symanzik equations \cite{Callan:1970yg}. However, dim reg has significant shortcomings. By analytically continuing only the spacetime indices—while keeping fields in representations of
$\mathrm{O}(1,3)$—dim reg breaks supersymmetry (SUSY) \cite{Siegel:1979wq} and unitarity \cite{Hogervorst:2015akt}. Additionally, the Clifford algebra 
$\{\gamma^\mu,\gamma^\nu\}=\eta^{\mu\nu}\mathbb{I}$ is assumed to hold for the $4\times4$ $\gamma$-matrices, even though the metric satisfies 
$\eta^{\mu\nu}\eta_{\mu\nu}=d$, leading to inconsistencies between trace and contraction identities. The definition of the $\gamma^5$
matrix also becomes ambiguous. The Breitenlohner–Maison–'t Hooft–Veltman (BMHV) prescription resolves this ambiguity by introducing an infinite number of $\gamma$
matrices, but at the cost of explicitly breaking Lorentz invariance \cite{tHooft:1972tcz,de_Mello_Koch_2020}. Furthermore, the Levi-Civita symbol cannot be consistently defined, complicating the calculation of the axial anomaly \cite{Peskin:1995ev,Novotny:1994yx}. Extending dim reg to spacetimes that are not maximally symmetric is also problematic, as dim reg relies on rotational symmetry and flat metrics \cite{Hawking:1976ja}, making it ill-suited for general finite-volume or compactified spacetimes. Moreover, dim reg can even yield incorrect results for certain finite integrals \cite{Treiman:1986ep}. Most fundamentally, dim reg is an \emph{ad hoc} method that is far from rigorous: one performs non-trivial manipulations of formally divergent integrals (e.g., shifting integration variables) when evaluating the Dyson series \emph{before} one renders these integrals finite via dim reg, thus casting significant doubt on any final result.
\par Dimensional reduction \cite{Siegel:1979wq} modifies dim reg such that only loop momenta are treated in $d$ dimensions, while all field and $\gamma$ matrix indices are treated in 4 dimensions. As a result, dimensional reduction preserves SUSY; however, ambiguity in defining and using $\gamma^5$ remains. Consequently, dimensional reduction can yield an incorrect axial anomaly or break SUSY and gauge invariance at one-loop order \cite{Stockinger:2005gx}.

\par Zeta function regularization was originally developed as an unambiguous way to handle divergences in curved spacetimes \cite{Hawking:1976ja}. Its higher-loop extension is known as operator regularization \cite{McKeon:1986rc}. However, both approaches generally violate BRS symmetry \cite{Rebhan:1988ed}.

\par Denominator regularization (den reg) is a more recent proposal that attempts to combine the strengths of these methods while avoiding their drawbacks \cite{Horowitz:2022uak}. Here, loop integrals are regulated by combining denominators using Feynman parameters and analytically continuing the overall exponent from 
$n\to n+\epsilon$, with $\epsilon$ chosen to ensure convergence. A fictitious scale is introduced, and Green’s functions obey the Callan–Symanzik equations. Since the number of spacetime dimensions remains fixed, den reg preserves SUSY, avoids ambiguities in defining 
$\gamma^5$, and correctly reproduces the axial anomaly. Moreover, den reg applies to curved spacetimes, thermal field theory, and finite-sized systems \cite{DuPlessis:2023vju}. However, it remains unclear whether gauge and BRST invariance are preserved, and, like dim reg, the theory is not finite at the level of the Lagrangian; as a result, formally divergent quantities are manipulated, potentially leading to non-unique results.  

\par Analytic regularization (analytic reg) generalizes propagator exponents to complex parameters 
$\lambda_i$, where each propagator in a loop carries its own analytic continuation \cite{speer}. This method yields Green's functions that are analytic in 
$\lambda_i$, with simple poles at $\lambda_i=1$, and is consistent with the renormalization schemes of Bogoliubov, Parasiuk, and Hepp \cite{speer}. The method builds upon analytic techniques developed by Riesz and Hadamard to preserve Lorentz invariance and causality\footnote{Despite locality being broken, the Green's functions remain causal \cite{Bollini1964}.}\cite{Bollini1964}. Like den reg, analytic reg maintains a fixed number of spacetime dimensions, preserving SUSY at leading order and providing an unambiguous definition of $\gamma^5$. Much like den reg, this approach results in an analytically continued power of the denominator, ensuring convergence of the integrals. Analytic reg also naturally incorporates a scale $\mu$, ensuring compatibility with RG flow via the Callan–Symanzik equations. Although slightly more involved than dim reg due to additional terms introduced through Feynman parameterization, these terms are finite and can be managed through suitable renormalization. Naïve applications to QED and Yang–Mills theories in three spacetime dimensions preserve gauge invariance, but in two and four spacetime dimensions, gauge symmetry is broken \cite{Hand:1990em,Manzoni:1998hp}. Furthermore, in \cite{Heydeman:2020ijz} a more rigorous approach to applying analytic reg was employed by analytically continuing the power of the kinetic term of the photon at the level of the action, while leaving the structure of the electron unchanged. The modification utilized in \cite{Heydeman:2020ijz} introduced non-local terms to the action. It was shown that for integer spacetime dimension $d\le3$, gauge invariance is not broken, however, for $d\ge4$; the theory is divergent in such a way that the regulator does not tame the divergences, and the regularization scheme fails. Several \emph{ad hoc} methods have been proposed to restore gauge invariance in analytic regularization. One approach introduces multi-photon vertices to enforce the Ward–Takahashi identity, thereby preserving gauge invariance \cite{Kroll1966}. Another modifies the polynomial structure of the electron momentum so that divergent integrals, which would otherwise violate gauge invariance, become convergent \cite{Breitenlohner:1977hr}. While changing the structure of the electron propagator can be successful—for example, in reproducing the axial anomaly in 2D QED \cite{Manzoni:1998hp}—these changes are purpose-built to preserve gauge invariance without due consideration of how such modifications may fundamentally affect the underlying quantum field theory. Additionally, the method introduced in \cite{Breitenlohner:1977hr} modifies the structure of the electron propagator while leaving the structure of the photon propagator unchanged, which breaks SUSY \cite{Gates:1983nr}.
\subsection{A New Approach}   
\par We seek a regularization scheme that is mathematically rigorous, symmetry-preserving, and conceptually well-defined. Building on the provisional strategy of \cite{Kroll1966}, in which the vertices of QED are modified in order to maintain Ward identities, we propose a more systematic method: Systematic Analytic Regularization (SAR). The approach introduced by SAR generalizes analytic regularization at the level of the action by analytically continuing the powers of kinetic differential operators. SAR preserves both finiteness and symmetry at the level of the quantum field theory (QFT) action, prior to the computation of Feynman diagrams. This approach is expected to yield more mathematically consistent results, with gauge invariance manifestly preserved. The systematic nature of SAR stands in contrast to other regularization schemes, which are applied at the level of Feynman diagrams and involve heuristic manipulations of formally divergent quantities.
\par SAR yields Lorentz-invariant, non-local kinetic terms that correspond to well-defined propagators. This non-locality suggests a conceptual connection to string theory, where point-like interactions are naturally smeared. SAR retains a fixed number of spacetime dimensions, avoids ambiguities in defining $\gamma^5,$ and ensures compatibility with the Callan–Symanzik equations and (presumably) preserves gauge invariance and SUSY; confirming the preservation of gauge invariance and SUSY will be the subject of future work.
\par To test SAR, we apply SAR to $\varphi^4$ and Yukawa theories at NLO and demonstrate that SAR successfully regularizes these theories while remaining relatively simple to implement. That is, SAR is capable of rendering all divergences in the respective theories finite at NLO, up to a potential pole in the regulator. This pole can then be canceled through the process of renormalization, ensuring that all final results are independent of the regulator and therefore comparable to experiments.
\par It is important to note that analytically continuing the kinetic terms of the action has been applied in the study of other areas of physics, primarily in the field of condensed matter (e.g. \cite{Fisher:1972zz,Benedetti:2020rrq}). While using a similar approach to SAR, the physics being explored is vastly different. In both \cite{Fisher:1972zz} and \cite{Benedetti:2020rrq}, long range scalar models are being studied in order to determine the critical exponents and fixed points of the renormalization group. The methods used in \cite{Fisher:1972zz} and \cite{Benedetti:2020rrq} fix the spacetime dimensions to be less than 4 and the exponents of the kinetic operators to be less than 1, which is in stark contrast to our approach in which we fix the dimension of spacetime to be 4 and the exponent of the kinetic operators to be greater than 1. Moreover, in our work, we introduce a renormalization scale and a regulator dependent phase factor $e^{-i\frac{\pi\epsilon}{2}}$ that preserves unitarity and is absent in \cite{Fisher:1972zz} and \cite{Benedetti:2020rrq}. Both \cite{Fisher:1972zz} and \cite{Benedetti:2020rrq} limit their studies to various scalar field theories; in our work we study both scalars and fermions with the ultimate goal of including gauge fields.      
\par Finally, a rather detailed and thorough treatment of analytically continued kinetic operators can be found in \cite{Calcagni:2021ljs}. Our work differs from \cite{Calcagni:2021ljs} in several important ways. First, \cite{Calcagni:2021ljs} studied the unitarity of theories with analytically continued kinetic operators and concluded that in order for the theory to remain unitary, the power of the kinetic term should be less than 1. In our approach, we introduce a crucial phase that depends on the regulator, which guarantees that theories with kinetic operators with powers greater than 1 are unitary.  Secondly, \cite{Calcagni:2021ljs} strictly studied theories involving only scalar fields, whereas we study a scalar theory and a theory with both scalars and fermions.  Finally, unlike \cite{Calcagni:2021ljs}, our work introduces a renormalization scale at the level of the action, which is important for RG considerations such as computing anomalous dimensions, running couplings, etc.

\section{The Superficial Degree of Divergence}
In computing amplitudes from Feynman diagrams, we often encounter divergences. In order to study and classify the ultraviolet (UV) divergences of a theory, we introduce the notion of the superficial degree of divergence. We will restrict ourselves to massive theories; hence we will not have to worry about infrared (IR) divergences. The superficial degree of divergence of a diagram is defined as the power of the loop
momentum in the numerator minus the power of the loop momentum in the denominator. If the superficial degree of divergence of a diagram is greater than or equal to zero, then the diagram has a possible UV divergence and is called superficially divergent. A theorem by Weinberg \cite{Weinberg} states that if a diagram and all its subgraphs have a negative superficial degree of divergence, then the diagram is convergent.
\par Let us recall some terminology. A \emph{subdiagram} is a subset of lines and vertices from a diagram that itself forms a diagram constructed using the same set of Feynman rules as the original. A diagram is said to be \emph{connected} if it consists of a single component connected in the topological sense. A \emph{truncated} diagram is a Green's function where the external propagators have been removed. A \emph{one-particle irreducible} (1PI) diagram is a truncated diagram that cannot be separated into two disjoint subdiagrams by cutting a single line. Any diagram of a theory can be written as the sum of products of 1PI diagrams \cite{Sterman:1993hfp}. Hence finding the relevant divergences in a theory reduces to finding the divergent 1PI diagrams, which means that the relevant superficial degree of divergence of a theory is the superficial degree of divergence of the 1PI diagrams of that theory. Therefore, the successful regularization of a theory is indicated by rendering all 1PI diagrams of a theory finite. In order to show that SAR is an effective regularization scheme, we will investigate all divergent 1PI diagrams at NLO in $\varphi^4$ and Yukawa theory.
\section{$\varphi^4$ Theory}
We first consider $\varphi^4$ theory. The bare action of $\varphi^4$ theory is given by
\begin{equation}\label{eq:phi^4bareaction}
    S_{\varphi^4}\equiv\int d^4x\left(\frac{1}{2}\varphi_0(-\square-M_0^2+i\varepsilon)\varphi_0-\frac{\lambda_0}{4!}\varphi_0^4\right),
\end{equation}
where $\varphi_0$ is the bare scalar field, $M_0$ is the bare mass, $\lambda_0$ is the bare coupling constant, and $\varepsilon$ is a small, positive number that shifts the poles in the $p^0$ plane, implementing time-ordering. We begin with a classical analysis of \eqref{eq:phi^4bareaction}. We consider all the symmetries of the theory:
\begin{itemize}
    \item The action \eqref{eq:phi^4bareaction} is invariant under the full Poincaré group.
    \item The action \eqref{eq:phi^4bareaction} is invariant under the finite group $\mathbb{Z}_2$ that acts on $\varphi_0$ as 
    $\varphi_0\mapsto-\varphi_0$.
\end{itemize}
\par From \eqref{eq:phi^4bareaction}, we can read off the Feynman rules of the bare theory:
\begin{equation}\label{eq:phi4bareFR}
    \begin{split}
        &\begin{tikzpicture}
        \draw[dashed] (1,0)--(-1,0);
        \node[above] (0,0) {$p$};
        \node[right] at (1,0){$\;=\frac{i}{p^2-M_0^2+i\varepsilon}$,};
        \end{tikzpicture}\\
    &\begin{tikzpicture}
        \filldraw[black] (0,0) circle (0.05);
        \draw[dashed] (1,-1)--(0,0);
        \draw[dashed] (1,1)--(0,0);
        \draw[dashed] (-1,-1)--(0,0);
        \draw[dashed] (-1,1)--(0,0);
        \node[right] at (1,0) {$\;=-i\lambda_0$.};
    \end{tikzpicture}
    \end{split}
\end{equation}
\par
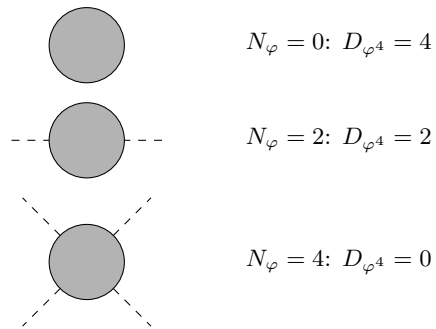
\begin{figure}[h]
        \centering
        \begin{align*}
        \begin{tikzpicture}
            \filldraw[color=black, fill=black!30] (0,0) circle (0.5);
            \node[right] at (2,0) {$N_\varphi=0$: $D_{\varphi^4}=4$};
        \end{tikzpicture}\\
        \begin{tikzpicture}
            \filldraw[color=black, fill=black!30] (0,0) circle (0.5);
            \draw[dashed] (-1,0)--(-0.5,0);
            \draw[dashed] (1,0)--(0.5,0);
            \node[right] at (2,0) {$N_\varphi=2$: $D_{\varphi^4}=2$};
        \end{tikzpicture}\\
        \begin{tikzpicture}
            \filldraw[color=black, fill=black!30] (0,0) circle (0.5);
            \draw[dashed] (0.353553,0.353553)--(0.853553,0.853553);
            \draw[dashed] (-0.353553,0.353553)--(-0.853553,0.853553);
            \draw[dashed] (0.353553,-0.353553)--(0.853553,-0.853553);
            \draw[dashed] (-0.353553,-0.353553)--(-0.853553,-0.853553);
            \node[right] at (2,0) {$N_\varphi=4$: $D_{\varphi^4}=0$};
        \end{tikzpicture}\\
        \end{align*}
        \caption{The superficially divergent 1PI diagrams of $\varphi^4$ theory.}
        \label{fig:phi4sdg}
    \end{figure}

 We would like to find all the potentially divergent 1PI diagrams of $\varphi^4$ theory, since these diagrams will need to be regularized. In order to find the potentially divergent diagrams, we compute the superficial degree of divergence of all the possible diagrams in $\varphi^4$ theory. Let us introduce the following definitions:
\begin{subequations}
    \begin{align}
        L\equiv&\;\text{the number of loop integrals},\\
        P_\varphi\equiv&\;\text{the number of internal propagators},\\
        N_\varphi\equiv&\;\text{the number of external propagators},\\
        V\equiv&\;\text{the number of vertices}.
    \end{align}
\end{subequations}
Let us notice that each loop integral introduces a factor of $d^4p\sim p^4$, while each internal propagator introduces a factor that scales as $\sim p^{-2}$. Note that external legs do not contribute to the superficial degree of divergence of 1PI diagrams. Hence, the superficial degree of divergence of a diagram in the theory, $D_{\varphi^4}$, is given by

\begin{equation}\label{eq:phi4defssd}
    D_{\varphi^4}\equiv 4L-2P_\varphi.
\end{equation}
Now, we should notice that we can relate the number of loop integrals to the number of internal propagators and vertices. The number of loop integrals is equal to the number of unconstrained momenta in a diagram. Each internal propagator comes with an associated momentum, which increases the number of unconstrained momenta by 1. Each vertex in a diagram introduces a delta function that imposes momentum conservation; however, one of these delta functions imposes overall energy-momentum conservation. Hence, the number of loop integrals is given by  
\begin{equation}\label{eq:phi4LinPV}
    L=P_\varphi-V+1.
\end{equation}
Next, from \eqref{eq:phi4bareFR}, there are 4 lines meeting at each vertex. Each external propagator has to be attached to a vertex at one end, while each internal propagator is attached to a vertex at both ends. Hence we have
\begin{equation}\label{eq:phi4VinNP}
    4V=N_\varphi+2P_\varphi.
\end{equation}
We can then solve \eqref{eq:phi4VinNP} to write $P_\varphi$ in terms of $V$ and $N_\varphi$ to find
\begin{equation}\label{eq:phi4PinVN}
    P_\varphi=2V-\frac{1}{2}N_\varphi.
\end{equation}
Then substituting both \eqref{eq:phi4LinPV} and \eqref{eq:phi4VinNP} into \eqref{eq:phi4defssd}, we find
\begin{equation}\label{eq:phi4unregssd}
    D_{\varphi^4}=4-N_\varphi.
\end{equation}
From \eqref{eq:phi4unregssd}, the 1PI diagrams with $N_\varphi=0,1,2,3,4$ are all superficially divergent. However due to the $\mathbb{Z}_2$-symmetry of the theory under $\varphi\to-\varphi$, only diagrams with an even number of external legs are non-vanishing\footnote{An alternative explanation is that for any graph with an odd number of external legs, there is no way to fully Wick contract all fields in the time ordered expansion of the interaction picture time-translation operator.}. So the superficially divergent diagrams are those with $N_\varphi=0,2,4$, which are depicted in \cref{fig:phi4sdg}.
\par In order to cancel the divergences of the theory, we introduce the renormalized field $\varphi_r$, the renormalized mass $M_r$ and the renormalized coupling $\lambda_r$ that are connected to the bare quantities by the renormalization constants $Z_\varphi$, $Z_M$, $Z_\lambda$ as follows\footnote{Note that we are simply relabeling the quantities in the original action. The divergences only appear when we compute Feynman diagrams.}:
\begin{subequations}\label{eq:phi4renormquantities}
    \begin{align}
        \varphi_r\equiv&\,Z_\varphi^{-\frac{1}{2}}\varphi_0\\
        M_r^2\equiv&\,Z_M^{-1}M_0^2\\
        \lambda_r\equiv&\,Z_\lambda^{-1}\lambda_0.
    \end{align}
\end{subequations}
We can then rewrite \eqref{eq:phi^4bareaction} in terms of the renormalized quantities \eqref{eq:phi4renormquantities} as
\begin{equation}\label{eq:phi4renormcnstaction}
    S_{\varphi^4}=\int d^4x\left(\frac{Z_\varphi}{2}\varphi_r(-\square-Z_MM_r^2+i\varepsilon)\varphi_r-\frac{Z_\lambda Z_\varphi^2\lambda_r}{4!}\varphi_r^4\right).
\end{equation}
Next, we define the shifts
\begin{subequations}\label{eq:phi4shiftsdef}
    \begin{align}
        \delta_\varphi\equiv&\,Z_\varphi-1\\
        \delta_M\equiv&\,M_r^2(Z_\varphi Z_M-1)\\
        \delta_\lambda\equiv&\,\lambda_r(Z_\lambda Z_\varphi^2-1).
    \end{align}
\end{subequations}
In terms of the shifts \eqref{eq:phi4shiftsdef}, we can rewrite \eqref{eq:phi4renormcnstaction} as
\begin{equation}\label{eq:phi4actionshifts}
\begin{split}
    S_{\varphi^4}=&\int d^4x\biggl(\frac{1}{2}\varphi_r(-\square-M_r^2+i\varepsilon)\varphi_r-\frac{\lambda_r}{4!}\varphi_r^4\\
    &+\frac{1}{2}\varphi_r(-\delta_\varphi\square-\delta_M)\varphi_r-\frac{\delta_\lambda}{4!}\varphi_r^4\biggr).
\end{split}
\end{equation}
\subsection{Systematic Analytic Regularization}
Let us define the regularized action as
\begin{equation}\label{eq:phi4fracregaction}
    \begin{split}
        S_{\text{SAR},\epsilon}\equiv&\int d^4x\biggl(\frac{\mu^{-\epsilon}e^{-\frac{i\pi\epsilon}{2}}}{2}\varphi_r(-\square-M_r^2+i\varepsilon)^{1+\frac{\epsilon}{2}}\varphi_r\\&-\frac{\lambda_r}{4!}\varphi_r^4
        +\frac{1}{2}\varphi_r(-\delta_\varphi\square-\delta_M)\varphi_r-\frac{\delta_\lambda}{4!}\varphi_r^4\biggr),
    \end{split}
\end{equation}
where $\epsilon>0$ is introduced as a UV-regulator; $\mu$ is an arbitrary constant with dimensions of mass, referred to as the renormalization scale; and the phase $e^{-\frac{i\pi\epsilon}{2}}$ \cite{speer} is required by unitarity. The fractional differential operator $(-\square-M_r^2+i\varepsilon)^{1+\frac{\epsilon}{2}}$ is defined by its Fourier transform
\begin{equation}\begin{split}\label{eq:deffracderiv}
    &(-\square-M_r^2+i\varepsilon)^{1+\frac{\epsilon}{2}}f(x)\\
    &\equiv\int\frac{d^4p}{(2\pi)^4}(p^2-M_r^2+i\varepsilon)^{1+\frac{\epsilon}{2}}\tilde{f}(p)e^{-ip\cdot x}\,,
    \end{split}
\end{equation}
where $f\in C^{\infty}$ and decays ``fast enough" at infinity. In \eqref{eq:deffracderiv}, $\tilde{f}(p)$ is the Fourier transform of $f(x)$,
\begin{equation}
    \tilde{f}(p)\equiv\int d^4x f(x)e^{ip\cdot x}\,.
\end{equation}
Hence, the kinetic term of \eqref{eq:phi4fracregaction} is given by
\begin{equation}\label{eq:phi4frakkindef}
    S_{\text{kin},\epsilon}\equiv\int\frac{d^4p}{(2\pi)^4}\frac{(\mu e^{\frac{i\pi}{2}})^{-\epsilon}}{2}\tilde{\varphi}_r(-p)(p^2-M_r^2+i\varepsilon)^{1+\frac{\epsilon}{2}}\tilde{\varphi}_r(p).
\end{equation}
Since the action \eqref{eq:phi4frakkindef} is defined in terms of Fourier space, rather than local fields $\varphi(x)$ and derivatives $\square$, the theory is non-local.
\par Notice that
\begin{equation}
    S_{\varphi^4}=\lim_{\epsilon\to0}S_{\text{SAR},\epsilon},
\end{equation}
so that when we smoothly remove the regulator, $\epsilon$, we recover the original $\varphi^4$ theory.
\par Let us now consider the symmetries of the fractionally regularized theory \eqref{eq:phi4fracregaction}:
\begin{itemize}
    \item The action \eqref{eq:phi4fracregaction} is invariant under the full Poincaré group.
    \item The action \eqref{eq:phi4fracregaction} is invariant under the finite group $\mathbb{Z}_2$ that acts on $\varphi_r$ as 
    $\varphi_r\mapsto-\varphi_r$.
\end{itemize}
Hence, SAR preserves all of the symmetries of the original theory \eqref{eq:phi^4bareaction}.
\par Let us now compute the superficial degree of divergence of \eqref{eq:phi4fracregaction}. Each loop integral introduces a factor of $d^4p\sim p^4$, while each scalar propagator scales as $\sim p^{-2-\epsilon}$, so we have that the superficial degree of divergence, $D_{\varphi^4,\text{SAR}}$, of the SAR theory \eqref{eq:phi4fracregaction} is given by the form
\begin{equation}\label{eq:ssdphi4fracstart}
    D_{\varphi^4,\;\text{SAR}}=4L-(2+\epsilon)P_\varphi.
\end{equation}
Both equations \eqref{eq:phi4LinPV} and \eqref{eq:phi4PinVN} still hold, which leads to
\begin{equation}\label{eq:ssdphi4frac}
    D_{\varphi^4,\;\text{SAR}}=4-N_\varphi-\epsilon\left(2V-\frac{1}{2}N_\varphi\right).
\end{equation}
Notice that as $\epsilon\to0$, we have that $ D_{\varphi^4,\text{SAR}}\to D_{\varphi^4}$ as expected. Since $\epsilon>0$ and, from \eqref{eq:phi4VinNP}, $2V\ge\frac{1}{2}N_\varphi$, we have that
\begin{equation}
     D_{\varphi^4,\;\text{SAR}}\leq D_{\varphi^4}; 
\end{equation}
thus, SAR has reduced the superficial degree of divergence of the theory.
\par The regularized Feynman rules can be read off from \eqref{eq:phi4fracregaction}:
\begin{equation}\label{eq:phi4fracregFR}
    \begin{split}
    &\begin{tikzpicture}
        \draw[dashed] (1,0)--(-1,0);
        \draw[->] (0.1,0)--(0,0);
        \node[above] (0,0) {$p$};
        \node[right] at (1,0){$=\frac{i\mu^\epsilon e^{\frac{i\pi\epsilon}{2}}}{(p^2-M_r^2+i\varepsilon)^{1+\frac{\epsilon}{2}}}$};
        \end{tikzpicture}\\
        &\begin{tikzpicture}
        \draw[dashed] (1,0)--(0.15,0);
        \draw[dashed] (-0.15,0)--(-1,0);
        \draw (0,0) circle (0.15);
        \draw (-0.106066,-0.106066)--(0.106066,0.106066);
        \draw (-0.106066,0.106066)--(0.106066,-0.106066);
        \node[right] at (1,0) {$=i(p^2\delta_\varphi-\delta_M)$};
        \node[above] at (0,0.15) {$p$};
    \end{tikzpicture}\\
    &\begin{tikzpicture}
        \filldraw[black] (0,0) circle (0.05);
        \draw[dashed] (1,-1)--(0,0);
        \draw[dashed] (1,1)--(0,0);
        \draw[dashed] (-1,-1)--(0,0);
        \draw[dashed] (-1,1)--(0,0);
        \node[right] at (1,0) {$=-i\lambda_r$};
    \end{tikzpicture}\\
    &\begin{tikzpicture}
        \draw[dashed] (1,-1)--(0,0);
        \draw[dashed] (1,1)--(0,0);
        \draw[dashed] (-1,-1)--(0,0);
        \draw[dashed] (-1,1)--(0,0);
        \draw (-0.106066,-0.106066)--(0.106066,0.106066);
        \draw (-0.106066,0.106066)--(0.106066,-0.106066);
        \draw (0,0) circle (0.15);
        \node[right] at (1,0) {$=-i\delta_{\lambda}.$};
    \end{tikzpicture}
    \end{split}
\end{equation}
\par We can now return to the superficially divergent diagrams in the unregularized theory given in \cref{fig:phi4sdg}. Let $-i\mathbf{V}_{(N_\varphi)}(\{p_i\})$ denote the $N_\varphi$-point 1PI diagram.
\par First, we notice that the zero-point diagram ($N_\varphi=0$) only contributes to an overall shift in the vacuum energy and, hence, cannot be measured. We will therefore ignore the zero point diagram.
\par
\begin{figure}[t]
    \centering
        \begin{tikzpicture}[scale=0.6,baseline={(0,0)}]
            \node[left,scale=0.9] at (-1,0) {$-i\pi(p^2)\equiv$};
            \draw[dashed] (-1,0)--(-0.5,0);
            \filldraw[color=black, fill=gray!20] (0,0) circle (0.5);
            \node at (0,0) {1PI};
            \draw[dashed] (0.5,0)--(1,0);
            \node[right,scale=0.8] at (1,0) {$=$};
        \end{tikzpicture}
        \begin{tikzpicture}[scale=0.6,baseline={(0,0)}]
            \draw[dashed] (-1,0)--(1,0);
            \draw[dashed] (0,0.5) circle (0.5);
            \filldraw[black] (0,0) circle (0.05);
            \node[right,scale=0.8] at (1,0) {$+$};
        \end{tikzpicture}
        \begin{tikzpicture}[scale=0.6,baseline={(0,0)}]
            \draw[dashed] (-1,0)--(1,0);
            \draw[dashed] (0,0) circle (0.5);
            \filldraw[black] (-0.5,0) circle (0.05);
            \filldraw[black] (0.5,0) circle (0.05);
            \node[right,scale=0.8] at (1,0) {$+\ldots+$};
        \end{tikzpicture}
        \begin{tikzpicture}[scale=0.6,baseline={(0,0)}]
            \draw[dashed] (-1,0)--(-0.15,0);
        \draw[dashed] (0.15,0)--(1,0);
        \draw (0,0) circle (0.15);
        \draw (-0.106066,-0.106066)--(0.106066,0.106066);
        \draw (-0.106066,0.106066)--(0.106066,-0.106066);
        \end{tikzpicture}
    \caption{The sum of all scalar 1PI two point diagrams in $\varphi^4$ theory.}
    \label{fig:phi4pi}
\end{figure}
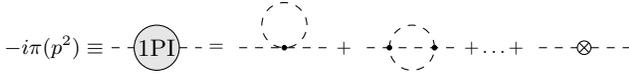

 Let us turn our attention to the two-point diagram ($N_\varphi=2$). Let $-i\pi(p^2)$ denote the sum of all 1PI scalar two point diagrams as depicted in \cref{fig:phi4pi}. We will refer to $\pi(p^2)$ as the scalar self-energy. 
Let us define the following one-loop diagram:
\begin{equation}\label{eq:defpi_s}
    \begin{tikzpicture}
            \node[left] at (-1,0) {$-i\pi_\text{s}^{(1)}(p^2)\equiv$};
            \draw[dashed] (-1,0)--(1,0);
            \draw[dashed] (0,0.5) circle (0.5);
            \filldraw[black] (0,0) circle (0.05);
            \draw[->] (0.05,1)--(-0.05,1);
            \node[above] at (0,1) {$k$};
            \node[right] at (1,0) {,};
        \end{tikzpicture}
\end{equation}
where the subscript $s$ is to indicate that there is a scalar loop, which will become important in Yukawa theory, where a fermion loop also contributes to the scalar self-energy. Then, at NLO, the scalar self-energy is given by
\begin{equation}
    -i\pi^{(1)}(p^2)=-i\pi^{(1)}_s(p^2)+i\left(p^2\delta_\varphi-\delta_M\right).
\end{equation}
In SAR \eqref{eq:defpi_s} is given by
\begin{equation}\label{eq:fracregpi_sfiniteeps}
    \begin{split}
        -i\pi^{(1)}_s(p^2)=&-\frac{i\lambda_r\mu^\epsilon}{2}\int\frac{d^4k}{(2\pi)^4}\frac{ie^{\frac{i\pi\epsilon}{2}}}{(k^2-M_r^2+i\varepsilon)^{1+\frac{\epsilon}{2}}}\\
        =&\frac{i\lambda_r\mu^\epsilon }{32\pi^2}\frac{\Gamma\left(\frac{\epsilon}{2}-1\right)}{\Gamma\left(1+\frac{\epsilon}{2}\right)}\left(M_r^2-i\varepsilon\right)^{1-\frac{\epsilon}{2}}.
    \end{split}
\end{equation}
Expanding \eqref{eq:fracregpi_sfiniteeps} about $\epsilon=0$ gives
\begin{equation}\label{eq:fracregpi_sinfiniteeps}
    \begin{split}
        -i\pi^{(1)}_s(p^2)=&-\frac{i\lambda_rM_r^2}{32\pi^2}\biggl[\frac{2}{\epsilon}-\gamma_E+1+\log\left(\frac{\mu^2}{M_r^2}\right)\biggr]\\
        &+\mathcal{O}\left(\epsilon\right),
    \end{split}
\end{equation}
where, since $M_r^2>0$, we have safely taken $\varepsilon\to0$.
\par Now, we are required to choose a set of counter-terms $\delta_\varphi$ and $\delta_M$ to cancel the $\epsilon^{-1}$ pole in \eqref{eq:fracregpi_sinfiniteeps}. Throughout this paper, we will use a slightly modified $\overline{\text{MS}}$ renormalization scheme where, since the number of dimensions is fixed, we do not need to subtract any $\log(4\pi)$'s. Much like in dim reg, we still subtract off the Euler-Mascheroni constant, $\gamma_E$. Since there is no dependence on $p^2$, we set
\begin{equation}
    \delta_\varphi=0.
\end{equation}
To cancel the $\epsilon^{-1}$ term and the Euler-Mascheroni constant, we set
\begin{equation}
    \delta_M=\frac{\lambda_r}{32\pi^2}\left(\frac{2}{\epsilon}-\gamma_E\right).
\end{equation}
The NLO correction to the scalar self-energy is then given by
\begin{equation}\label{eq:phi4finalpi1}
    \begin{split}
        \pi^{(1)}(p^2)=\frac{\lambda_rM_r^2}{32\pi^2}\left[1+\log\left(\frac{\mu^2}{M_r^2}\right)\right].
    \end{split}
\end{equation}
\par
\begin{figure}
    \centering
    \begin{tikzpicture}
            \node[left] at (-1,0) {$-i\mathbf{V}_{(4)}(\{p_i\})\equiv$};
            \filldraw[color=black, fill=gray!20] (0,0) circle (0.5);
            \node at (0,0) {1PI};
            \node at (1,1) {$p_4$};
            \node at (1,-1) {$p_2$};
            \node at (-1,-1) {$p_1$};
            \node at (-1,1) {$p_3$};
            \draw[->] (0.353553,0.353553)--(0.603553,0.603553);
            \draw[->] (-0.353553,0.353553)--(-0.603553,0.603553);
            \draw[->] (0.853553,-0.853553)--(0.603553,-0.603553);
            \draw[->] (-0.853553,-0.853553)--(-0.603553,-0.603553);
            \draw (0.353553,0.353553)--(0.853553,0.853553);
            \draw (-0.353553,0.353553)--(-0.853553,0.853553);
            \draw (0.353553,-0.353553)--(0.853553,-0.853553);
            \draw (-0.353553,-0.353553)--(-0.853553,-0.853553);
        \end{tikzpicture}
    \caption{The 4-point 1PI diagram.}
    \label{fig:phi441PI}
\end{figure}
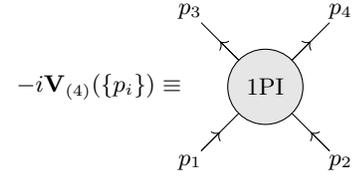

 Next, we consider the four point diagram ($N_\varphi=4$). Hence we will be computing $\mathbf{V}_{(4)}(\{p_i\})$ as depicted in \cref{fig:phi441PI}.
Let us define the following one-loop diagram $-iV(p^2)$:
\begin{equation}\label{eq:defV(p^2)}
    \begin{tikzpicture}
        \node[left] at (-1,0) {$-iV(p^2)\equiv$};
        \draw[dashed] (0,0) circle (0.5);
        \draw[dashed] (0,0.5)--(0.5,1);
        \filldraw[black] (0,-0.5) circle (0.05);
        \filldraw[black] (0,0.5) circle (0.05);
        \draw[dashed] (0,0.5)--(-0.5,1);
        \draw[dashed] (0,-0.5)--(0.5,-1);
        \draw[dashed] (0,-0.5)--(-0.5,-1);
        \draw[->] (0,-1)--(0,-0.6);
        \node[below] at (0,-1) {$p$};
        \draw[->] (-0.5,0.1)--(-0.5,0);
        \node[left] at (-0.5,0) {$k$};
        \draw[->] (0.5,0)--(0.5,0.1); 
        \node[right] at (0.5,0) {$p+k\;.$};
    \end{tikzpicture}
\end{equation}
Define the Mandelstam variables as per usual as 
\begin{subequations}
\label{eq:Mandelstam}
    \begin{align}
        s\equiv&\,(p_1+p_2)^2=(p_3+p_4)^2\\
        t\equiv&\,(p_3-p_1)^2=(p_4-p_2)^2\\
        u\equiv&\,(p_3-p_2)^2=(p_4-p_1)^2.
    \end{align}
\end{subequations}
Now, at next-to-leading order, we have that the four point vertex correction is given by
\begin{equation}\label{eq:phi44pointvertex}
    -i\mathbf{V}^{(1)}_{(4)}(\{p_i\})=-i\lambda_r-i\left(V(s)+V(t)+V(u)\right)-i\delta_\lambda.
\end{equation}
In SAR, $-iV(p^2)$, as defined in \eqref{eq:defV(p^2)}, is
\begin{equation}\label{eq:V(p^2)fracreg.step1}
    \begin{split}
        &-iV(p^2)=\frac{(-i\lambda_r\mu^\epsilon)^2}{2}\int\frac{d^4k}{(2\pi)^4}\frac{ie^{\frac{i\pi\epsilon}{2}}}{\left(k^2-M_r^2+i\varepsilon\right)^{1+\frac{\epsilon}{2}}}\\
        \times&\frac{ie^{\frac{i\pi\epsilon}{2}}}{\left((p+k)^2-M_r^2+i\varepsilon\right)^{1+\frac{\epsilon}{2}}}\\
        =&-\frac{i(\lambda_r\mu^\epsilon)^2}{2}\frac{\Gamma\left(2+\epsilon\right)}{\Gamma\left(1+\frac{\epsilon}{2}\right)^2}\int_0^1dxx^{\frac{\epsilon}{2}}(1-x)^{\frac{\epsilon}{2}}\\
        \times&\int\frac{d^4k}{(2\pi)^4}\frac{ie^{i\pi\epsilon}}{\left(k^2+2xk\cdot p+xp^2-M_r^2+i\varepsilon\right)^{2+\epsilon}}.
    \end{split}
\end{equation}
Define
\begin{equation}\label{eq:V(p^2)fracreg.lsub}
    \begin{split}
        &l^\mu\equiv k^\mu+xp^\mu,\\
        \implies&l^2-x(x-1)p^2=k^2+2xk\cdot p+xp^2,
    \end{split}
\end{equation}
and
\begin{equation}\label{eq:V(p^2)fracreg.Delta}
    \Delta^2\equiv M_r^2-x(1-x)p^2.
\end{equation}
Then with the $l$-substitution \eqref{eq:V(p^2)fracreg.lsub} and $\Delta^2$ as given in \eqref{eq:V(p^2)fracreg.Delta}, we can write \eqref{eq:V(p^2)fracreg.step1} as
\begin{equation}\label{eq:V(p^2)fracreg.finiteeps}
    \begin{split}
        &-iV(p^2)=-\frac{i(\lambda_r\mu^\epsilon)^2}{2}\frac{\Gamma\left(2+\epsilon\right)}{\Gamma\left(1+\frac{\epsilon}{2}\right)^2}\\
        &\times\int_0^1dxx^{\frac{\epsilon}{2}}(1-x)^{\frac{\epsilon}{2}}\int\frac{d^4l}{(2\pi)^4}\frac{i}{\left(l^2-\Delta^2+i\varepsilon\right)^{2+\epsilon}}\\
        =&-\frac{i(\lambda_r\mu^\epsilon)^2}{32\pi^2\Gamma\left(1+\frac{\epsilon}{2}\right)^2}\\
        &\times\int_0^1dx\frac{x^{\frac{\epsilon}{2}}(1-x)^{\frac{\epsilon}{2}}\Gamma\left(\epsilon\right)}{\left(M_r^2-x(1-x)p^2-i\varepsilon\right)^{\epsilon}}.
    \end{split}
\end{equation}
Taking the $\epsilon\to0$ limit of \eqref{eq:V(p^2)fracreg.finiteeps} gives
\begin{equation}\label{eq:V(p^2)fracreg.infiniteep}
    \begin{split}
        -iV(p^2)=&-\frac{i\lambda_r^2}{32\pi^2}\int_0^1dx\biggl[\frac{1}{\epsilon}-\gamma_E-2\\
        &+\log\left(\frac{\mu^2}{M_r^2-x(x-1)p^2-i\varepsilon}\right)\biggr]+\mathcal{O}(\epsilon).
    \end{split}
\end{equation}
To cancel the $\epsilon^{-1}$ term and the Euler-Mascheroni constant, we choose the shift to be of the form:
\begin{equation}
    \delta_\lambda=-\frac{3\lambda_r^2}{32\pi^2}\biggl(\frac{1}{\epsilon}-\gamma_E\biggr).
\end{equation}
The NLO corrected four-point vertex \eqref{eq:phi44pointvertex} is then given by
\begin{equation}\label{eq:NLOV4phi4}
    \begin{split}
        &-i\mathbf{V}^{(1)}_{(4)}(\{p_i\})=-i\lambda_r\\
        &-\frac{i\lambda_r^2}{32\pi^2}\int_0^1dx\biggl(\log\left(\frac{\mu^2}{M_r^2-x(1-x)s-i\varepsilon}\right)-6\\
        &+\log\left(\frac{\mu^2}{M_r^2-x(1-x)t}\right)+\log\left(\frac{\mu^2}{M_r^2-x(1-x)u}\right)\biggr),
    \end{split}
\end{equation}
where, since $t,u\le0$, we can take the $\varepsilon\to0$ limit for the last two terms. Notice that, had we not included the factor of $e^{-\frac{i\pi\epsilon}{2}}$ in the action \eqref{eq:phi4fracregaction}, the $\lambda_r^2$ term in \eqref{eq:NLOV4phi4} would contain a factor of $(-1)^{\epsilon}=e^{i\pi\epsilon}=1+i\pi\epsilon+\mathcal{O}(\epsilon^2)$. The imaginary part of \eqref{eq:NLOV4phi4} would then be changed. As a result, the Optical Theorem \cite{Peskin:1995ev} would no longer be satisfied, and thus the theory would violate unitarity. 
\par Through computing the superficial degree of divergence of $\varphi^4$ theory, we identified all the superficially divergent diagrams of the theory. We then showed, to NLO, that SAR self-consistently and effectively regularizes $\varphi^4$ theory, yielding finite results for all the superficially divergent diagrams of the theory at NLO. We can compare \eqref{eq:phi4finalpi1} and \eqref{eq:NLOV4phi4} to known textbook results (e.g. \cite{SkinnerQFTII,Peskin:1995ev}) and see that they agree up to renormalization scheme dependent constant terms such as $\gamma_E$ and $\log4\pi$.
\section{Yukawa Theory}
We now consider Yukawa theory. The bare action of the Yukawa theory is described by the bare Dirac bispinor $\psi_0$, with bare mass $m_0$ and a real bare scalar field $\varphi_0$ with bare mass $M_0$ and bare couplings $g_0$ and $\lambda_0$:
\begin{equation}\label{eq:bareYukawaaction}
    \begin{split}
        S_\text{Yukawa}\equiv&\int d^4x\biggl(\bar{\psi}_0(i\slashed{\p}-m_0+i\varepsilon)\psi_0\\
        &+\frac{1}{2}\varphi_0(-\square-M_0^2+i\varepsilon)\varphi_0\\
        &-g_0\varphi_0\psib_0\psi_0-\frac{\lambda_0}{4!}\varphi_0^4\biggr),
    \end{split}
\end{equation}
where $\psib\equiv\psi^\dagger\gamma^0$, $\slashed{\p}\equiv\gamma^\mu\p_\mu$, and $\varepsilon$ is a small positive number that shifts the poles in the $p^0$ plane, implementing time-ordering.
\par Let us consider all the symmetries of the theory \eqref{eq:bareYukawaaction}: 
\begin{itemize}
    \item The Yukawa theory is invariant under the full Poincaré group.
    \item The Yukawa action is invariant under the global $\mathrm{U}(1)$ symmetry that acts as follows
    \begin{subequations}\label{eq:globalU(1)}
        \begin{align}
            \varphi_0&\mapsto\varphi_0\\
            \psi_0&\mapsto e^{i\theta}\psi_0\\
            \psib_0&\mapsto e^{-i\theta}\psib_0,
        \end{align}
    \end{subequations}
    where $\theta\in\mathbb{R}$.
    \item The Yukawa action is invariant under the global $\mathbb{Z}_2$ symmetry that acts as follows
    \begin{subequations}\label{eq:Z2Yukawa}
        \begin{align}
            \varphi_0&\mapsto\varphi_0\\
            \psi_0&\mapsto -\psi_0\\
            \psib_0&\mapsto -\psib_0\,.
        \end{align}
    \end{subequations}
    Notice that the $\mathbb{Z}_2$ symmetry is actually just a finite subgroup of the $\mathrm{U}(1)$ symmetry, where $\theta=\pi$.
\end{itemize}
\par From \eqref{eq:bareYukawaaction}, we can read off the bare Feynman rules:
\begin{equation}\label{eq:bareFeynmanYukawa}
    \begin{split}
        &\begin{tikzpicture}
        \draw (1,0)--(-1,0);
        \draw[->] (0.1,0)--(0,0);
        \node[above] (0,0) {$p$};
        \node[right] at (1,0){$=\frac{i(\slashed{p}+m_0)}{p^2-m_0^2+i\varepsilon}$};
        \end{tikzpicture}\\
    &\begin{tikzpicture}
        \draw[dashed] (1,0)--(-1,0);
        \draw[->] (0.1,0)--(0,0);
        \node[above] (0,0) {$p$};
        \node[right] at (1,0){$=\frac{i}{p^2-M_0^2+i\varepsilon}$};
        \end{tikzpicture}\\
    &\begin{tikzpicture}
    \filldraw[black] (0,0) circle (0.05);
     \draw[dashed] (-1,0)--(0,0);
     \draw (0,0)--(0.5,0.866);
     \draw[->] (0.25,0.433)--(0.275,0.476);
     \draw (0,0)--(0.5,-0.866);
     \draw[->] (0.25,-0.476)--(0.225,-0.379);
     \node[right] at (1,0) {$=-ig_0$};
    \end{tikzpicture}\\
    &\begin{tikzpicture}
        \filldraw[black] (0,0) circle (0.05);
        \draw[dashed] (1,-1)--(0,0);
        \draw[dashed] (1,1)--(0,0);
        \draw[dashed] (-1,-1)--(0,0);
        \draw[dashed] (-1,1)--(0,0);
        \node[right] at (1,0) {$=-i\lambda_0.$};
    \end{tikzpicture}
    \end{split}
\end{equation}
\par
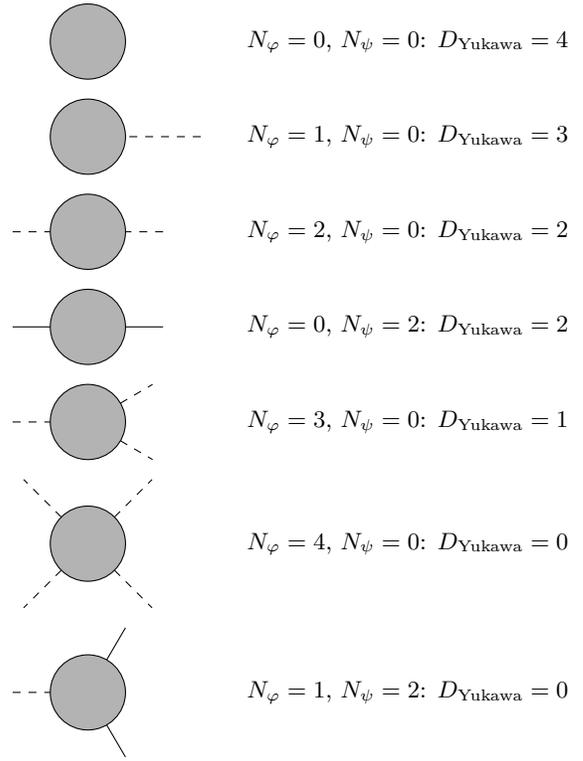
\begin{figure}[h]
    \centering
    \begin{align*}
    \begin{tikzpicture}
        \filldraw[color=black, fill=black!30] (0,0) circle (0.5);
        \node[right] at (2,0) {$N_\varphi=0$, $N_\psi=0$: $D_\text{Yukawa}=4$};
    \end{tikzpicture}\\
    \begin{tikzpicture}
        \filldraw[color=black, fill=black!30] (0,0) circle (0.5);
        \draw[dashed] (1.5,0)--(0.5,0);
        \node[right] at (2,0) {$N_\varphi=1$, $N_\psi=0$: $D_\text{Yukawa}=3$};
    \end{tikzpicture}\\
    \begin{tikzpicture}
        \filldraw[color=black, fill=black!30] (0,0) circle (0.5);
        \draw[dashed] (-1,0)--(-0.5,0);
        \draw[dashed] (1,0)--(0.5,0);
        \node[right] at (2,0) {$N_\varphi=2$, $N_\psi=0$: $D_\text{Yukawa}=2$};
    \end{tikzpicture}\\
    \begin{tikzpicture}
        \filldraw[color=black, fill=black!30] (0,0) circle (0.5);
        \draw (-1,0)--(-0.5,0);
        \draw (1,0)--(0.5,0);
        \node[right] at (2,0) {$N_\varphi=0$, $N_\psi=2$: $D_\text{Yukawa}=2$};
    \end{tikzpicture}\\
    \begin{tikzpicture}
        \filldraw[color=black, fill=black!30] (0,0) circle (0.5);
        \draw[dashed] (-1,0)--(-0.5,0);
        \draw[dashed] (0.433,0.25)--(0.86,0.5);
        \draw[dashed] (0.433,-0.25)--(0.86,-0.5);
        \node[right] at (2,0) {$N_\varphi=3$, $N_\psi=0$: $D_\text{Yukawa}=1$};
    \end{tikzpicture}\\
    \begin{tikzpicture}
        \filldraw[color=black, fill=black!30] (0,0) circle (0.5);
        \draw[dashed] (0.353553,0.353553)--(0.853553,0.853553);
        \draw[dashed] (-0.353553,0.353553)--(-0.853553,0.853553);
        \draw[dashed] (0.353553,-0.353553)--(0.853553,-0.853553);
        \draw[dashed] (-0.353553,-0.353553)--(-0.853553,-0.853553);
        \node[right] at (2,0) {$N_\varphi=4$, $N_\psi=0$: $D_\text{Yukawa}=0$};
    \end{tikzpicture}\\
    \begin{tikzpicture}
        \filldraw[color=black, fill=black!30] (0,0) circle (0.5);
        \draw[dashed] (-1,0)--(-0.5,0);
        \draw (0.25,0.433)--(0.5,0.86);
        \draw (0.25,-0.433)--(0.5,-0.86);
        \node[right] at (2,0) {$N_\varphi=1$, $N_\psi=2$: $D_\text{Yukawa}=0$};
    \end{tikzpicture}
    \end{align*}
    \caption{The divergent irreducible diagrams of Yukawa theory.}
    \label{fig:sdgYukawa}
\end{figure}

 Next, we consider the superficial degree of divergence of the theory. Let us introduce the following definitions:
\begin{subequations}
    \begin{align}
        &L\equiv\;\text{number of loop integrals},\\
        &P_\psi\equiv\;\text{number of internal fermion propagators},\\
        &P_\varphi\equiv\;\text{number of internal scalar propagators},\\
        &N_\psi\equiv\;\text{number of external fermion propagators},\\
        &N_\varphi\equiv\;\text{number of external scalar propagators},\\
        &V\equiv\;\text{number of vertices}.
    \end{align}
\end{subequations}
We would like to find and regularize all possible divergences of the theory, as was done in the previous section, we consider the superficial degree of divergence of the theory, $D_\text{Yukawa}$. Each loop integral introduces a measure $d^4p\sim p^4$, while the scalar propagators of the theory scales as $\sim p^{-2}$ and the fermionic propagators scales as $\sim p^{-1}$. Hence, the form of $D_\text{Yukawa}$ is given by
\begin{equation}\label{eq:defssdyukawa}
    D_\text{Yukawa}\equiv 4L-2P_\varphi-P_\psi\,.
\end{equation}
The number of loop integrals is equal to the number of unconstrained momenta in a diagram. Each internal propagator comes with an associated momentum, which increases the number of unconstrained momenta by 1. Each vertex in a diagram introduces a delta function that imposes energy-momentum conservation; however, one of these delta functions imposes overall energy-momentum conservation. Hence, the number of loop integrals is given by
\begin{equation}\label{eq:LintermspsiphiV}
    L=P_\psi+P_\varphi-V+1,
\end{equation}
Each internal scalar line must connect to two vertices, and each external scalar line connects to a single vertex, which gives the following relation
\begin{equation}\label{eq:Vintermsphi}
    V=2P_\varphi+N_\varphi\,,
\end{equation}
and for each scalar at a vertex, there must be two fermions, so we also have 
\begin{equation}\label{eq:Vintermspsi}
    V=\frac{1}{2}(2P_\psi+N_\psi).
\end{equation}
From  \eqref{eq:Vintermsphi} and \eqref{eq:Vintermspsi}, we can write
\begin{subequations}\label{eq:internalpsiphiinVexternal}
    \begin{align}
        P_\varphi=&\frac{1}{2}(V-N_\varphi)\\
        P_\psi=&V-\frac{1}{2}N_\psi.
    \end{align}
\end{subequations}
From \eqref{eq:internalpsiphiinVexternal}, we can write \eqref{eq:LintermspsiphiV} as
\begin{equation}\label{eq:LYukawaexternal}
    L=\frac{1}{2}(V-N_\varphi-N_\psi)+1.
\end{equation}
Then, from \eqref{eq:internalpsiphiinVexternal} and \eqref{eq:LYukawaexternal}, we can write \eqref{eq:defssdyukawa} in terms of external propagators and vertices as
\begin{equation}\label{eq:NonregssdYukawa}
    D_\text{Yukawa}=4-N_\varphi-\frac{3}{2}N_\psi,
\end{equation}
From \eqref{eq:NonregssdYukawa}, we show the superficially divergent diagrams in \cref{fig:sdgYukawa}.
Note that, from the $\mathbb{Z}_2$ symmetry \eqref{eq:Z2Yukawa}, all diagrams with an odd number of external fermion lines are identically 0.
\par In order to cancel the divergences of the theory, we renormalize the theory by introducing the renormalized fields $\psi_r$ and $\varphi_r$, renormalized masses $m_r$ and $M_r$, and renormalized couplings $g_r$ and $\lambda_r$ that are connected to the bare quantities by the renormalization constants $Z_\psi$, $Z_\varphi$, $Z_m$, $Z_M$, $Z_g$ and $Z_\lambda$ as follows:
\begin{subequations}\label{eq:renormquanY}
    \begin{align}
        \psi_r\equiv&\, Z_\psi^{-\frac{1}{2}}\psi_0\\
        \varphi_r\equiv&\,Z_\varphi^{-\frac{1}{2}}\varphi_0\\
        m_r\equiv&\, Z_m^{-1}m_0\\
        M_r^2\equiv&\,Z_M^{-1}M_0^2\\
        g_r\equiv&\,Z_g^{-1}g_0\\
        \lambda_r\equiv&\,Z_\lambda^{-1}\lambda_0.
    \end{align}
\end{subequations}
We can then rewrite \eqref{eq:bareYukawaaction} in terms of the renormalized quantities \eqref{eq:renormquanY} as 
\begin{equation}\label{eq:renormYukawacnst}
    \begin{split}
        S_\text{Yukawa}=&\int d^4x\biggl(Z_\psi\oll{\psi}_r(i\slashed{\p}-Z_mm_r+i\varepsilon)\psi_r\\
        &+\frac{Z_\varphi}{2}\varphi_r(-\square-Z_MM_r^2+i\varepsilon)\varphi_r\\
        &-Z_gZ_\psi Z^{\frac{1}{2}}_\varphi g_r\varphi_r\psib_r\psi_r-\frac{Z_\lambda Z_\varphi^2\lambda_r}{4!}\varphi_r^4\biggr).
    \end{split}
\end{equation}
Define the shifts 
\begin{subequations}\label{eq:shiftsYukawa}
    \begin{align}
        \delta_\psi\equiv&\, Z_\psi-1\\
        \delta_\varphi\equiv&\, Z_\varphi-1\\
        \delta_m\equiv&\,m_r\left(Z_\psi Z_m-1\right)\\
        \delta_M\equiv&\,M_r^2\left(Z_\varphi Z_M-1\right)\\
        \delta_g\equiv&\,g_r\left(Z_gZ_\psi Z_\varphi^\frac{1}{2}-1\right)\\
        \delta_\lambda\equiv&\,\lambda_r\left(Z_\lambda Z_\varphi^2-1\right).
    \end{align}
\end{subequations}
In terms of the shifts \eqref{eq:shiftsYukawa}, we can write the action \eqref{eq:renormYukawacnst} as
\begin{equation}\label{eq:renormYukawashifts}
\begin{split}
    S_\text{Yukawa}
=&\int d^4x\biggl(
\psib_r(i\slashed{\partial}-m_r+i\varepsilon)\psi_r\\
&
+\frac{1}{2}\varphi_r(-\square-M_r^2+i\varepsilon)\varphi_r\\
&-g_r\varphi_r\psib_r\psi_r-\frac{\lambda_r}{4!}\varphi_r^4\\
&+\psib_r(i\delta_\psi\slashed{\partial}-\delta_m)\psi_r\\
&+\frac{1}{2}\varphi_r(-\delta_\varphi\square-\delta_M)\varphi_r\\
&-\delta_g\varphi_r\psib_r\psi_r
-\frac{\delta_\lambda}{4!}\varphi_r^4
\biggr).
\end{split}
\end{equation}
Using the Clifford algebra, we can write the following formal expression for the fermionic kinetic operator:
\begin{equation}\label{eq:fermionickineticterm}
    (i\slashed{\p}-m_r+i\varepsilon)=\frac{(-\square-m_r^2+i\varepsilon)}{i\slashed{\p}+m_r}\,.
\end{equation}
Note that, had we multiplied by a factor of $(i\slashed{\p}+m_r-i\varepsilon)/(i\slashed{\p}+m_r-i\varepsilon)$, we would get a purely real quantity in the denominator of the propagator, which affects the time-ordering properties of the propagator.   
Thus, we can rewrite \eqref{eq:renormYukawashifts} as
\begin{equation}\label{eq:renormYukawashiftsfermkinetic}
  \begin{split}
S_\text{Yukawa}=&\int d^4x\biggl(
\psib_r\frac{-\square-m_r^2+i\varepsilon}{i\slashed{\partial}+m_r}\psi_r\\
&-g_r\varphi_r\psib_r\psi_r
+\frac{1}{2}\varphi_r(-\square-M_r^2+i\varepsilon)\varphi_r\\
&-\frac{\lambda_r}{4!}\varphi_r^4
+\psib_r(i\delta_\psi\slashed{\partial}-\delta_m)\psi_r\\
&+\frac{1}{2}\varphi_r(-\delta_\varphi\square-\delta_M)\varphi_r\\
&-\delta_g\varphi_r\psib_r\psi_r
-\frac{\delta_\lambda}{4!}\varphi_r^4
\biggr).
\end{split}
\end{equation}

\subsection{Systematic Analytic Regularization}
We define the fractionally regularized action
\begin{equation}\label{eq:fracregYukawa}
    \begin{split}
        S_{\text{SAR},\epsilon}
=&\int d^4x\biggl(
\mu^{-\epsilon}e^{-\frac{i\pi\epsilon}{2}}\psib_r
\frac{
(-\square-m_r^2+i\varepsilon)^{1+\frac{\epsilon}{2}}}
{(i\slashed{\partial}+m_r)}\psi_r\\
&+\frac{\mu^{-\epsilon}e^{-\frac{i\pi\epsilon}{2}}}{2}
\varphi_r(-\square-M_r^2+i\varepsilon)^{1+\frac{\epsilon}{2}}\varphi_r\\
&-g_r\varphi_r\psib_r\psi_r
-\frac{\lambda_r}{4!}\varphi_r^4\\
&+\psib_r(i\delta_\psi\slashed{\partial}-\delta_m)\psi_r\\
&+\frac{1}{2}\varphi_r(-\delta_\varphi\square-\delta_M)\varphi_r\\
&-\delta_g\varphi_r\psib_r\psi_r
-\frac{\delta_\lambda}{4!}\varphi_r^4
\biggr),
    \end{split}
\end{equation}
where $\epsilon>0$ is a UV-regulator; $\mu$ is an arbitrary constant with dimensions of mass, referred to as the renormalization scale; and, again, the phase $e^{-\frac{i\pi\epsilon}{2}}$ is required by unitarity \cite{speer}. Now notice that 
\begin{equation}
    S_\text{Yukawa}=\lim_{\epsilon\to0}S_{\text{SAR},\epsilon}\,,
\end{equation}
that is, as we remove the regulator $\epsilon$, we recover the original Yukawa theory.
\par Let us consider the symmetries of the systematically regularized theory \eqref{eq:fracregYukawa}:
\begin{itemize}
    \item The action \eqref{eq:fracregYukawa} is invariant under the full Poincaré group.
    \item The action \eqref{eq:fracregYukawa} is invariant under the global $\mathrm{U}(1)$ symmetry \eqref{eq:globalU(1)}.
    \item The action \eqref{eq:fracregYukawa} is invariant under the $\mathbb{Z}_2$ symmetry \eqref{eq:Z2Yukawa}.
\end{itemize}
Hence, all the original symmetries of the theory \eqref{eq:bareYukawaaction} are preserved, as desired.
\par Let us now compute the superficial degree of divergence of the SAR theory \eqref{eq:fracregYukawa}. Each loop introduces a factor of $d^4p\sim p^4$, each scalar propagator scales as $\sim p^{-2-\epsilon}$ and each fermion propagator scales as $\sim p^{-1-\epsilon}$. Therefore, the superficial degree of divergence, $D_{\text{Yukawa, SAR}}$, of the theory \eqref{eq:fracregYukawa} is given by:
\begin{equation}\label{eq:ssdyukawafracstart}
    D_{\text{Yukawa, SAR}}=4L-(2+\epsilon)P_\varphi-(1+\epsilon)P_\psi.
\end{equation}
Equations \eqref{eq:internalpsiphiinVexternal} and \eqref{eq:LYukawaexternal} still hold, so we find that the superficial degree of divergence of the SAR theory \eqref{eq:fracregYukawa} can be written as
\begin{equation}\label{eq:ssdyukawafrac}\begin{split}
    D_{\text{Yukawa, SAR}}=&4-N_\varphi-\frac{3}{2}N_\psi\\&-\epsilon\left(\frac{3}{2}V-\half N_\varphi-N_\psi\right).
\end{split}\end{equation}
Notice that as $\epsilon\to0$, we have $D_{\text{Yukawa, SAR}}\to D_{\text{Yukawa}}$, as expected. Furthermore, since $\epsilon>0$ and, from \eqref{eq:Vintermsphi} and \eqref{eq:Vintermspsi}, $3V\ge N_\varphi+2N_\psi$, we have that 
\begin{equation}
    D_{\text{Yukawa, SAR}}\le D_{\text{Yukawa}};
\end{equation}
thus, SAR has reduced the superficial degree of divergence of the theory.
\par From \eqref{eq:fracregYukawa}, we can read off the regularized Feynman rules:
\begin{equation}
    \begin{split}
        &\begin{tikzpicture}
        \draw (1,0)--(-1,0);
        \draw[->] (0.1,0)--(0,0);
        \node[above] (0,0) {$p$};
        \node[right] at (1,0){$=\frac{i\mu^\epsilon e^{\frac{i\pi\epsilon}{2}}(\slashed{p}+m_r)}{(p^2-m_r^2+i\varepsilon)^{1+\frac{\epsilon}{2}}}$};
        \end{tikzpicture}\\
        &\begin{tikzpicture}
        \draw (1,0)--(0.15,0);
        \draw (-0.15,0)--(-1,0);
        \draw (0,0) circle (0.15);
        \draw (-0.106066,-0.106066)--(0.106066,0.106066);
        \draw (-0.106066,0.106066)--(0.106066,-0.106066);
        \node[right] at (1,0) {$=i(\slashed{p}\delta_\psi-\delta_m)$};
        \node[above] at (0,0.15) {$p$};
    \end{tikzpicture}\\
    &\begin{tikzpicture}
        \draw[dashed] (1,0)--(-1,0);
        \node[above] (0,0) {$p$};
        \node[right] at (1,0){$=\frac{i\mu^{\epsilon}e^{\frac{i\pi\epsilon}{2}}}{(p^2-M_r^2+i\varepsilon)^{1+\frac{\epsilon}{2}}}$};
        \end{tikzpicture}\\
        &\begin{tikzpicture}
        \draw[dashed] (1,0)--(0.15,0);
        \draw[dashed] (-0.15,0)--(-1,0);
        \draw (0,0) circle (0.15);
        \draw (-0.106066,-0.106066)--(0.106066,0.106066);
        \draw (-0.106066,0.106066)--(0.106066,-0.106066);
        \node[right] at (1,0) {$=i(p^2\delta_\varphi-\delta_M)$};
        \node[above] at (0,0.15) {$p$};
    \end{tikzpicture}\\
    &\begin{tikzpicture}
    \filldraw[black] (0,0) circle (0.05);
     \draw[dashed] (-1,0)--(0,0);
     \draw (0,0)--(0.5,0.866);
     \draw[->] (0.25,0.433)--(0.275,0.476);
     \draw (0,0)--(0.5,-0.866);
     \draw[->] (0.25,-0.476)--(0.225,-0.379);
     \node[right] at (1,0) {$=-ig_r$};
    \end{tikzpicture}\\
    &\begin{tikzpicture}
    \draw (0,0) circle (0.15);
        \draw (-0.106066,-0.106066)--(0.106066,0.106066);
        \draw (-0.106066,0.106066)--(0.106066,-0.106066);
     \draw[dashed] (-1,0)--(-0.15,0);
     \draw (0.075,0.1299)--(0.5,0.866);
     \draw (0.075,-0.1299)--(0.5,-0.866);
     \draw[->] (0.25,0.433)--(0.275,0.476);
     \draw[->] (0.25,-0.476)--(0.225,-0.379);
     \node[right] at (1,0) {$=-i\delta_g$};
    \end{tikzpicture}\\
    &\begin{tikzpicture}
        \filldraw[black] (0,0) circle (0.05);
        \draw[dashed] (1,-1)--(0,0);
        \draw[dashed] (1,1)--(0,0);
        \draw[dashed] (-1,-1)--(0,0);
        \draw[dashed] (-1,1)--(0,0);
        \node[right] at (1,0) {$=-i\lambda_r$};
    \end{tikzpicture}\\
    &\begin{tikzpicture}
        \draw[dashed] (1,-1)--(0,0);
        \draw[dashed] (1,1)--(0,0);
        \draw[dashed] (-1,-1)--(0,0);
        \draw[dashed] (-1,1)--(0,0);
        \draw (-0.106066,-0.106066)--(0.106066,0.106066);
        \draw (-0.106066,0.106066)--(0.106066,-0.106066);
        \draw (0,0) circle (0.15);
        \node[right] at (1,0) {$=-i\delta_{\lambda}.$};
    \end{tikzpicture}
    \end{split}
\end{equation}
Each fermion loop comes with a minus sign and a trace over the fermion propagators in the loop.
\par Now that we have all the necessary tools, let us turn to the divergent diagrams \cref{fig:sdgYukawa}. Let $-i\mathbf{V}_{(N_\varphi,N_\psi)}(\{p_i\})$ denote the 1PI diagram with $N_\varphi$ external scalar lines and $N_\psi$ external fermion lines.
\par Notice that the zero point diagram (with $N_\varphi=N_\psi=0$) only contributes to the vacuum energy and, hence, cannot be measured, so we will ignore it.
\par Next, let us focus on the diagrams with an odd number of external scalars and no external fermions ($N_\varphi=1,3$ and $N_\psi=0$). The diagram that contributes to $N_\varphi=1$ and $N_\psi=0$ is at LO 
\begin{equation}\label{eq:Nphi=1,Npsi=0}
    \begin{tikzpicture}
        \draw[dashed] (-1,0)--(0,0);
        \filldraw[black] (-1,0) circle (0.05);
        \draw (-1.5,0) circle (0.5);
        \draw[->] (-2,0)--(-2,-0.05);
        \node[right] at (0,0) {$.$};
    \end{tikzpicture}
\end{equation}
Let us recall that \cite{Peskin:1995ev}
\begin{equation}\label{eq:anti-commutivity} \contraction{}{\psib_I}{(x)}{\psi_I}
\psib_I(x)\psi_I(y)
=
-
\contraction{}{\psib_I}{\psi_I}{(y)}
\psi_I(y)\bar\psi_I(x)
\end{equation}
Let us also note that the fermion loop implies that we must take a trace. Thus \eqref{eq:Nphi=1,Npsi=0} is given by
\begin{equation}\label{eq:V(1,0)equal-V(1,0)}
    \begin{split}
        \tr\bigl(\contraction{}{\psib_I}{(x)}{\psi_I}
\psib_I(x)\psi_I(x)\bigr)
        =&\tr\bigl(\contraction{}{\psib_I}{\psi_I}{(x)}
\psi_I(x)\psib_I(x)\bigr)\\
        =&-\tr\bigl(\contraction{}{\psib_I}{(x)}{\psi_I}
\psib_I(x)\psi_I(x)\bigr),
    \end{split}
\end{equation}
where in the first equality of \eqref{eq:V(1,0)equal-V(1,0)} we have used the cyclicity of the trace, and in the second equality we have used \eqref{eq:anti-commutivity}. From \eqref{eq:V(1,0)equal-V(1,0)}, we have that $\mathbf{V}^{(1)}_{(1,0)}(p^2)=-\mathbf{V}^{(1)}_{(1,0)}(p^2)$, and thus 
\begin{equation}
    \mathbf{V}^{(1)}_{(1,0)}(p^2)=0.
\end{equation}
\par The diagrams that contribute to $N_\varphi=3$ and $N_\psi=0$ at LO are 
\begin{equation}\label{eq:Nphi=3,Npsi=0}
    \begin{tikzpicture}
        \draw[dashed] (-1.5,0)--(-0.5,0);
        \draw[dashed] (0.25,0.433)--(0.75,1.299);
        \draw[dashed] (0.25,-0.433)--(0.75,-1.299);
        \filldraw[black] (-0.5,0) circle (0.05);
        \filldraw[black] (0.25,-0.433) circle (0.05);
        \filldraw[black] (0.25,0.433) circle (0.05);
        \draw (0,0) circle (0.5);
        \draw[->] (0.5,0)--(0.5,-0.05);
        \draw[->] (-0.25,-0.433)--(-0.255,-0.428);
        \draw[->] (-0.25,0.433)--(-0.245,0.438);
        \node[right] at (1,0) {$+$};
    \end{tikzpicture}
    \begin{tikzpicture}
        \draw[dashed] (-1.5,0)--(-0.5,0);
        \draw[dashed] (0.25,0.433)--(0.75,1.299);
        \draw[dashed] (0.25,-0.433)--(0.75,-1.299);
        \filldraw[black] (-0.5,0) circle (0.05);
        \filldraw[black] (0.25,-0.433) circle (0.05);
        \filldraw[black] (0.25,0.433) circle (0.05);
        \draw (0,0) circle (0.5);
        \draw[->] (0.5,-0.05)--(0.5,0);
        \draw[->] (-0.255,-0.428)--(-0.25,-0.433);
        \draw[->] (-0.245,0.438)--(-0.25,0.433);
        \node[right] at (1,0) {$.$};
    \end{tikzpicture}
\end{equation}
Note that, again, the fermion loops require us to take a trace. Let us define
\begin{equation}\begin{split}\label{eq:defD_F}
    D_F(x-y)\equiv&\contraction{}{\psib_I}{\psi_I}{(x)}
\psi_I(x)\psib_I(y)\\
    =&\bra{0}T(\psi_I(x)\psib_I(y))\ket{0}\,.
    \end{split}
\end{equation}
where $\ket{0}$ is the vacuum of the free theory, and $T$ denotes time ordering. Let us consider the action of charge conjugation $C$ (for details about charge conjugation and spinors, see, e.g., \cite{Srednicki:2007qs}). The action of $C$ on $\psi_I(x)$ and $\psib_I(x)$ is given by
\begin{equation}
    C\psi_I(x)C^{-1}= \mathcal{C}\psib^T(x)\quad\text{and}\quad C\psib_I(x)C^{-1}=\psi_I^T(x)\mathcal{C}\,,
\end{equation}
where $\mathcal{C}$ is the charge conjugation matrix \cite{Srednicki:2007qs}. Note that 
$\mathcal{C}$ satisfies 
\begin{equation}\label{eq:Cinv}
    \mathcal{C}^{-1}=-\mathcal{C}\,.
\end{equation}
Now we can rewrite \eqref{eq:defD_F} as follows
\begin{equation}\label{eq:derivationusingcc}
    \begin{split}
        D_F(x-y)
    =&\bra{0}T(CC^{-1}\psi_I(x)CC^{-1}\psib_I(y)CC^{-1})\ket{0}\\
    =&\bra{0}T(\psi_I^C(x)\psib_I^C(y))\ket{0}\\
    =&\,\mathcal{C}\bra{0}T(\psib^T_I(x)\psi_I^T(y))\ket{0}\mathcal{C}\\
    =&\,\mathcal{C}\bigl(\bra{0}T(\psi_I(y)\psib_I(x))\ket{0}\bigr)^T\mathcal{C}\,,
    \end{split}
\end{equation}
where we have assumed that the free theory vacuum $\ket{0}$, is invariant under $C$. Hence, we have that \eqref{eq:defD_F} satisfies 
\begin{equation}\label{eq:DFchargeconj}
    D_F(x-y)=\mathcal{C}D_F(y-x)^T\mathcal{C}\,.
\end{equation}
From \eqref{eq:DFchargeconj} and \eqref{eq:Cinv}, we can write propagators on the left-hand Feynman diagram of \eqref{eq:Nphi=3,Npsi=0} terms of \eqref{eq:defD_F} as follows
\begin{equation}\label{eq:Nphi=3,Npsi=0,Left}
    \begin{split}
        -\tr\bigl(D_F(x-y)D_F(y-z)D_F(z-x)\bigr)\,.
    \end{split}
\end{equation}
While the propagators on the right-hand Feynman diagram of \eqref{eq:Nphi=3,Npsi=0} takes the form
\begin{equation}\label{eq:Nphi=3,Npsi=0,Right}
    \begin{split}
        &-\tr\bigl(D_F(x-z)D_F(y-x)D_F(z-y)\bigr)\\=&-\tr\bigl(\mathcal{C}D_F(z-x)^{T}\mathcal{C}\mathcal{C}D_F(x-y)^{T}\mathcal{C}\mathcal{C}D_F(y-z)^{T}\mathcal{C}\bigr)\\
        =&-\tr\bigl(\mathcal{C}\mathcal{C}D_F(z-x)^{T}\mathcal{C}\mathcal{C}D_F(x-y)^{T}\mathcal{C}\mathcal{C}D_F(y-z)^{T}\bigr)\\
        =&(-1)^4\tr\bigl((D_F(x-y)D_F(y-z)D_F(z-x))^{T}\bigr)\\
        =&\tr\bigl(D_F(x-y)D_F(y-z)D_F(z-x)\bigr)\,.
    \end{split}
\end{equation}
From \eqref{eq:Nphi=3,Npsi=0,Left} and \eqref{eq:Nphi=3,Npsi=0,Right}, we have 
\begin{equation}\label{eq:Nphi3,Npsi=0,=0}
    \begin{tikzpicture}
        \draw[dashed] (-1.5,0)--(-0.5,0);
        \draw[dashed] (0.25,0.433)--(0.75,1.299);
        \draw[dashed] (0.25,-0.433)--(0.75,-1.299);
        \filldraw[black] (-0.5,0) circle (0.05);
        \filldraw[black] (0.25,-0.433) circle (0.05);
        \filldraw[black] (0.25,0.433) circle (0.05);
        \draw (0,0) circle (0.5);
        \draw[->] (0.5,0)--(0.5,-0.05);
        \draw[->] (-0.25,-0.433)--(-0.255,-0.428);
        \draw[->] (-0.25,0.433)--(-0.245,0.438);
        \node[right] at (1,0) {$=-$};
    \end{tikzpicture}
    \begin{tikzpicture}
        \draw[dashed] (-1.5,0)--(-0.5,0);
        \draw[dashed] (0.25,0.433)--(0.75,1.299);
        \draw[dashed] (0.25,-0.433)--(0.75,-1.299);
        \filldraw[black] (-0.5,0) circle (0.05);
        \filldraw[black] (0.25,-0.433) circle (0.05);
        \filldraw[black] (0.25,0.433) circle (0.05);
        \draw (0,0) circle (0.5);
        \draw[->] (0.5,-0.05)--(0.5,0);
        \draw[->] (-0.255,-0.428)--(-0.25,-0.433);
        \draw[->] (-0.245,0.438)--(-0.25,0.433);
        \node[right] at (1,0) {$.$};
    \end{tikzpicture}
\end{equation}
Hence, from \eqref{eq:Nphi=3,Npsi=0}, \eqref{eq:Nphi=3,Npsi=0}, and \eqref{eq:Nphi3,Npsi=0,=0}, we see that
\begin{equation}
    \mathbf{V}^{(1)}_{(3,0)}(\{p_i\})=0.
\end{equation}
\par Let us now turn to the two point diagrams. First, we consider the scalar two point function ($N_\varphi=2$, $N_\psi=0$). Let $-i\pi(p^2)$ denote the sum of all 1PI scalar two point diagrams as depicted in \cref{fig:piYukawa}. We will refer to $\pi(p^2)$ as the scalar self-energy.
\begin{figure}[t]
    \centering
        \begin{tikzpicture}[scale=0.6,baseline={(0,0)}]
            \node[left,scale=0.9] at (-1,0) {$-i\pi(p^2)\equiv$};
            \draw[dashed] (-1,0)--(-0.5,0);
            \filldraw[color=black, fill=gray!20] (0,0) circle (0.5);
            \node at (0,0) {1PI};
            \draw[dashed] (0.5,0)--(1,0);
            \node[right,scale=0.8] at (1,0) {$=$};
        \end{tikzpicture}
        \begin{tikzpicture}[scale=0.6,baseline={(0,0)}]
            \draw[dashed] (-1,0)--(1,0);
            \draw[dashed] (0,0.5) circle (0.5);
            \filldraw[black] (0,0) circle (0.05);
            \node[right,scale=0.8] at (1,0) {$+$};
        \end{tikzpicture}
        \begin{tikzpicture}[scale=0.6,baseline={(0,0)}]
            \draw[dashed] (-1,0)--(-0.5,0);
            \draw[dashed] (0.5,0)--(1,0);
            \draw (0,0) circle (0.5);
            \filldraw[black] (-0.5,0) circle (0.05);
            \filldraw[black] (0.5,0) circle (0.05);
            \node[right,scale=0.8] at (1,0) {$+\ldots+$};
        \end{tikzpicture}
        \begin{tikzpicture}[scale=0.6,baseline={(0,0)}]
            \draw[dashed] (-1,0)--(-0.15,0);
        \draw[dashed] (0.15,0)--(1,0);
        \draw (0,0) circle (0.15);
        \draw (-0.106066,-0.106066)--(0.106066,0.106066);
        \draw (-0.106066,0.106066)--(0.106066,-0.106066);
        \end{tikzpicture}
    \caption{The sum of all scalar 1PI two point diagrams in Yukawa theory.}
    \label{fig:piYukawa}
\end{figure}
Let us define the following one loop diagrams
\begin{equation}\label{eq:Yukawadefpi_fpi_s}
    \begin{split}
        &\begin{tikzpicture}
            \node[left] at (-1,0) {$-i\pi^{(1)}_\text{s}(p^2)\equiv$};
            \draw[dashed] (-1,0)--(1,0);
            \draw[dashed] (0,0.5) circle (0.5);
            \filldraw[black] (0,0) circle (0.05);
            \draw[->] (0.05,1)--(-0.05,1);
            \node[above] at (0,1) {$k$};
        \end{tikzpicture}\\
        &\begin{tikzpicture}
            \node[left] at (-1,0) {$-i\pi^{(1)}_\text{f}(p^2)\equiv$};
            \draw[dashed] (-1,0)--(-0.5,0);
            \draw[dashed] (0.5,0)--(1,0);
            \draw (0,0) circle (0.5);
            \filldraw[black] (-0.5,0) circle (0.05);
            \filldraw[black] (0.5,0) circle (0.05);
            \draw[->] (0.05,0.5)--(-0.05,0.5);
            \draw[->] (-0.05,-0.5)--(0.05,-0.5);
            \draw[->] (0.2,0.6).. controls (0.1,0.65) and (-0.1,0.65).. (-0.2,0.6);
            \draw[->] (-0.2,-0.6).. controls (-0.1,-0.65) and (0.1,-0.65).. (0.2,-0.6);
            \node[above] at (0,0.65) {$k$};
            \node[below] at (0,-0.65) {$p+k$};
        \end{tikzpicture}
    \end{split}
\end{equation}
Then, at NLO, the scalar self-energy is given by
\begin{equation}
    -i\pi^{(1)}(p^2)=-i\pi^{(1)}_\text{s}(p^2)-i\pi^{(1)}_\text{f}(p^2)+i(p^2\delta_\varphi-\delta_M).
\end{equation}
In SAR, $\pi_f(p^2)$, as defined in \eqref{eq:Yukawadefpi_fpi_s} becomes
\begin{equation}\label{eq:pi_ffracreg.step1}
    \begin{split}
        -i\pi^{(1)}_f(p^2)=&\,(g_r\mu^{\epsilon})^2\int\frac{d^4k}{(2\pi)^4}\tr\biggl(\frac{ie^{\frac{i\pi\epsilon}{2}}(\slashed{k}+m_r)}{(k^2-m_r^2+i\varepsilon)^{1+\frac{\epsilon}{2}}}\\
        &\times\frac{ie^{\frac{i\pi\epsilon}{2}}(\slashed{p}+\slashed{k}+m_r)}{((p+k)^2-m_r^2+i\varepsilon)^{1+\frac{\epsilon}{2}}}\biggr)\\
        =&\,4i(g_r\mu^{\epsilon})^2\frac{\Gamma\left(2+\epsilon\right)}{\Gamma\left(1+\frac{\epsilon}{2}\right)^2}\int_0^1dxx^{\frac{\epsilon}{2}}(1-x)^{\frac{\epsilon}{2}}\\
        &\times\int\frac{d^4k}{(2\pi)^4}\frac{ie^{i\pi\epsilon}(k^2+k\cdot p+m_r^2)}{D^{2+\epsilon}},
    \end{split}
\end{equation}
where
\begin{equation}
D\equiv k^2+2xk\cdot p+xp^2-m_r^2+i\varepsilon\,.
\end{equation}
Define
\begin{equation}\label{eq:pi_ffracreg.lsub}
    \begin{split}
        &l^{\mu}=k^\mu+xp^\mu\\
        \implies&k^2+2xk\cdot p+xp^2=l^2-x(x-1)p^2,
    \end{split}
\end{equation}
and define 
\begin{equation}\label{eq:pi_ffracreg.Delta}
    \Delta^2\equiv m_r^2-x(1-x)p^2.
\end{equation}
Then with the $l$-substitution \eqref{eq:pi_ffracreg.lsub} and $\Delta^2$ as given in \eqref{eq:pi_ffracreg.Delta}, we can write \eqref{eq:pi_ffracreg.step1} as 
\begin{equation}\label{eq:fracregpi_ffiniteeps}
    \begin{split}
        -i\pi^{(1)}_f(p^2)
        =&4i(g_r\mu^{\epsilon})^2\frac{\Gamma\left(2+\epsilon\right)}{\Gamma\left(1+\frac{\epsilon}{2}\right)^2}\int_0^1dxx^{\frac{\epsilon}{2}}(1-x)^{\frac{\epsilon}{2}}\\
        &\times\int\frac{d^4l}{(2\pi)^4}\frac{ie^{i\pi\epsilon}(l^2+\Delta^2)}{(l^2-\Delta^2+i\varepsilon)^{2+\epsilon}}\\
        =&\frac{i(g_r\mu^{\epsilon})^2(-1)^2}{4\pi^2\Gamma\left(1+\frac{\epsilon}{2}\right)^2}\int_0^1dxx^{\frac{\epsilon}{2}}(1-x)^{\frac{\epsilon}{2}}\\
        &\times\Delta^2(\Delta^2-i\varepsilon)^{-\epsilon}\biggl(\Gamma\left(\epsilon\right)-2\Gamma\left(-1+\epsilon\right)\biggr)\\
        =&\frac{i(g_r\mu^{\epsilon})^2}{4\pi^2\Gamma\left(1+\frac{\epsilon}{2}\right)^2}\int_0^1dxx^{\frac{\epsilon}{2}}(1-x)^{\frac{\epsilon}{2}}\\
        &\times\left(\Gamma\left(\epsilon\right)-2\Gamma\left(-1+\epsilon\right)\right)\\
        &\times\frac{m_r^2-x(1-x)p^2}{(m_r^2-x(1-x)p^2-i\varepsilon)^\epsilon}.
    \end{split}
\end{equation}
In the $\epsilon\ll1$ limit, we expand \eqref{eq:fracregpi_ffiniteeps} about $\epsilon=0$ to find
\begin{equation}\label{eq:pifinfiepsilon}
    \begin{split}
        -i\pi^{(1)}_f(p^2)=&-\frac{i3g_r^2}{4\pi^2}\int_0^1dx\left(x(1-x)p^2-m_r^2\right)\\
        &\times\biggl(\frac{1}{\epsilon}-\gamma_E+\log\left(\frac{\mu^2}{m_r^2-x(1-x)p^2-i\varepsilon}\right)\biggr)\\
        &+\mathcal{O}\left(\epsilon\right).
    \end{split}
\end{equation}
Then, by adding \eqref{eq:fracregpi_sinfiniteeps} and the corresponding counter terms, we find 
\begin{equation}
    \begin{split}
        \pi^{(1)}(p^2)=&\frac{\lambda_rM_r^2}{32\pi^2}\biggl(\frac{2}{\epsilon}-\gamma_E+1+\log\left(\frac{\mu^2}{M_r^2}\right)\biggr)\\
        &+\frac{3g_r^2}{4\pi^2}\int_0^1dx\left(x(1-x)p^2-m_r^2\right)\\
        &\times\left(\frac{1}{\epsilon}-\gamma_E+\log\left(\frac{\mu^2}{m_r^2-x(1-x)p^2-i\varepsilon}\right)\right)\\
        &+\delta_\varphi p^2-\delta_M.
    \end{split}
\end{equation}
To cancel the $p^2$-dependent $\epsilon^{-1}$ term and the Euler-Mascheroni constant, we set 
\begin{equation}
   \delta_\varphi=-\frac{g_r^2}{8\pi^2}\left(\frac{1}{\epsilon}-\gamma_E\right). 
\end{equation}
To cancel the $p^2$-independent $\epsilon^{-1}$ term and the Euler-Mascheroni constant, we set
\begin{equation}
    \delta_M=\frac{\lambda_rM_r^2}{32\pi^2}\left(\frac{2}{\epsilon}-\gamma_E\right)-\frac{3g_r^2m_r^2}{4\pi^2}\left(\frac{1}{\epsilon}-\gamma_E\right).
\end{equation}
The scalar self-energy, to NLO, is then given by:
\begin{equation}\label{eq:YukawaScalarSelfEnergy}
    \begin{split}
        \pi^{(1)}\left(p^2\right)=&\frac{\lambda_rM_r^2}{32\pi^2}\left(1+\log\left(\frac{\mu^2}{M_r^2}\right)\right)\\
        &+\frac{3g_r^2}{4\pi^2}\int_0^1dx\left(x(1-x)p^2-m_r^2\right)\\
        &\times\log\left(\frac{\mu^2}{m_r^2-x(1-x)p^2-i\varepsilon}\right).
    \end{split}
\end{equation}
\par Next, let us define $-i\Sigma(p^2)$ as the sum of all 1PI fermion two point diagrams as depicted in \cref{fig:SigmaYukawa}. We will refer to $\Sigma(p^2)$ as the fermion self-energy.
\begin{figure}[t]
    \centering
        \begin{tikzpicture}[scale=0.6,baseline={(0,0)}]
            \node[left,scale=0.9] at (-1,0) {$-i\Sigma(p^2)\equiv$};
            \draw (-1,0)--(-0.5,0);
            \filldraw[color=black, fill=gray!20] (0,0) circle (0.5);
            \node at (0,0) {1PI};
            \draw (0.5,0)--(1,0);
            \node[right,scale=0.8] at (1,0) {$=$};
        \end{tikzpicture}
        \begin{tikzpicture}[scale=0.6,baseline={(0,0)}]
            \draw (-1,0)--(1,0);
            \draw[dashed] (0.5,0) arc (0:180:0.5);
            \filldraw[black] (-0.5,0) circle (0.05);
            \filldraw[black] (0.5,0) circle (0.05);
            \node[right,scale=0.8] at (1,0) {$+$};
        \end{tikzpicture}
        \begin{tikzpicture}[scale=0.6,baseline={(0,0)}]
            \draw (-1,0)--(1,0);
            \draw[dashed] (0.2,0) arc (0:180:0.5); 
            \filldraw[black] (0.2,0) circle (0.05);
            \filldraw[black] (-0.8,0) circle (0.05);
            \draw[dashed] (0.7,0) arc (0:-180:0.5);
            \filldraw[black] (0.7,0) circle (0.05);
            \filldraw[black] (-0.3,0) circle (0.05);
            \node[right,scale=0.8] at (1,0) {$+\ldots+$};
        \end{tikzpicture}
        \begin{tikzpicture}[scale=0.6,baseline={(0,0)}]
            \draw (-1,0)--(-0.15,0);
        \draw (0.15,0)--(1,0);
        \draw (0,0) circle (0.15);
        \draw (-0.106066,-0.106066)--(0.106066,0.106066);
        \draw (-0.106066,0.106066)--(0.106066,-0.106066);
        \end{tikzpicture}
    \caption{The sum of all fermion 1PI two point diagrams.}
    \label{fig:SigmaYukawa}
\end{figure}
Define $-i\Sigma^{(1)}_s(p^2)$ as follows:
\begin{equation}\label{eq:YukawadefSigma}
    \begin{tikzpicture}
         \node[left] at (-1,0) {$-i\Sigma^{(1)}_s(p^2)\equiv$};
            \draw (-1,0)--(1,0);
            \draw[dashed] (0.5,0) arc (0:180:0.5);
            \filldraw[black] (-0.5,0) circle (0.05);
            \filldraw[black] (0.5,0) circle (0.05);
            \draw[->] (-0.05,0)--(0.05,0);
            \draw[->] (-0.2,0.6).. controls (0.1,0.65) and (-0.1,0.65).. (0.2,0.6);
            \draw[->] (0.2,-0.15)--(-0.2,-0.15);
            \node[above] at (0,0.65) {$p+k$};
            \node[below] at (0,-0.15) {$k$};
            \node[right] at (1,0) {$.$};
    \end{tikzpicture}
\end{equation}
Then, at NLO, the fermion self-energy is given by
\begin{equation}
    -i\Sigma^{(1)}(p^2)=-i\Sigma^{(1)}_s(p^2)+i(\slashed{p}\delta_\psi-\delta_m).
\end{equation}
Let us compute $-i\Sigma^{(1)}_s(p^2)$ as defined in \eqref{eq:YukawadefSigma} in SAR:
    \begin{equation}
    \begin{split}
        -i\Sigma^{(1)}_s(p^2)=&-(g_r\mu^{\epsilon})^2\int\frac{d^4k}{(2\pi)^4}\frac{e^{\frac{i\pi\epsilon}{2}}i(\slashed{k}+m_r)}{(k^2-m_r^2+i\varepsilon)^{1+\frac{\epsilon}{2}}}\\
        &\times\frac{ie^{\frac{i\pi\epsilon}{2}}}{((k+p)^2-M_r^2+i\varepsilon)^{1+\frac{\epsilon}{2}}}\\
        =&-i(g_r\mu^{\epsilon})^2\frac{\Gamma\left(2+\epsilon\right)}{\Gamma\left(1+\frac{\epsilon}{2}\right)^2}\int_0^1dxx^{\frac{\epsilon}{2}}(1-x)^{\frac{\epsilon}{2}}\\
        &\times\int\frac{d^4l}{(2\pi)^4}\frac{ie^{i\pi\epsilon}(\slashed{l}+x\slashed{p}+m_r)}{(l^2-\Delta^2+i\varepsilon)^{2+\epsilon}},
    \end{split}
\end{equation}
where 
\begin{equation}
    \begin{split}
        &l^\mu\equiv k^\mu+xp^\mu\\
        \implies&k^2+2xk\cdot p+xp^2=l^2-x(x-1)p^2,
    \end{split}
\end{equation}
and
\begin{equation}
    \Delta^2\equiv (1-x)m_r^2+xM_r^2-x(1-x)p^2.
\end{equation}
Then
\begin{equation}
    \begin{split}
        -i\Sigma^{(1)}_s(p^2)=&-i(g_r\mu^{\epsilon})^2\frac{\Gamma\left(2+\epsilon\right)}{\Gamma\left(1+\frac{\epsilon}{2}\right)^2}\int_0^1dxx^{\frac{\epsilon}{2}}(1-x)^{\frac{\epsilon}{2}}\\
        &\times\int\frac{d^4l}{(2\pi)^4}\frac{ie^{i\pi\epsilon}(x\slashed{p}+m_r)}{(l^2-\Delta^2+i\varepsilon)^{2+\epsilon}}\\
        =&-(-1)^{2}\frac{i(g_r\mu^{\epsilon})^2}{16\pi^2}\frac{\Gamma\left(\epsilon\right)}{\Gamma\left(1+\frac{\epsilon}{2}\right)^2}\\
        &\times\int_0^1dxx^{\frac{\epsilon}{2}}(1-x)^{\frac{\epsilon}{2}}(x\slashed{p}+m_r)\left(\Delta^2-i\varepsilon\right)^{\epsilon}\\
        =&\frac{i(g_r\mu^{\epsilon})^2}{16\pi^2}\frac{\Gamma\left(\epsilon\right)}{\Gamma\left(1+\frac{\epsilon}{2}\right)^2}\\
        &\times\int_0^1dxx^{\frac{\epsilon}{2}}(1-x)^{\frac{\epsilon}{2}}(x\slashed{p}+m_r)\left(\Delta^2-i\varepsilon\right)^{\epsilon}.
    \end{split}
\end{equation}
Now we can expand in $\epsilon$ to find 
\begin{equation}\label{eq:Sigmainfiniepsilon}
    \begin{split}
        &-i\Sigma^{(1)}_s(p^2)=-\frac{ig_r^2}{16\pi^2}\int_0^1dx(x\slashed{p}+m_r)
        \biggl(\frac{1}{\epsilon}-\gamma_E-1\\
        &+\log\left(\frac{\mu^2}{(1-x)m_r^2+xM_r^2-x(1-x)p^2-i\varepsilon}\right)\biggr).
    \end{split}
\end{equation}
To cancel the $\slashed{p}$-dependent $\epsilon^{-1}$ term and the Euler-Mascheroni constant, we set
\begin{equation}
  \delta_\psi=\frac{g_r^2}{16\pi^2}\left(\frac{1}{\epsilon}-\gamma_E\right).
\end{equation}
To cancel the $\slashed{p}$-independent $\epsilon^{-1}$ term and the Euler-Mascheroni constant, we set 
\begin{equation}
    \delta_m=-\frac{g_r^2m_r}{16\pi^2}\left(\frac{1}{\epsilon}-\gamma_E\right).
\end{equation}
Then, to NLO, the fermion self-energy is
\begin{equation}\label{eq:Yukawafermionselfenergy}
    \begin{split}
        &\Sigma^{(1)}(p^2)=\frac{ig_r^2}{16\pi^2}\int_0^1dx(x\slashed{p}+m_r)\\
        \times&\biggl[
        \log\left(\frac{\mu^2}{(1-x)m_r^2+xM_r^2-x(1-x)p^2-i\varepsilon}\right)-1\biggr].
    \end{split}
\end{equation}
\par Let us now consider the NLO contribution to the diagram with $N_\varphi=1$ and $N_\psi=2$. Define $\tilde{\mathbf{V}}_{(1,2)}(p,p')$ as given in \eqref{eq:defV_3}:
\begin{equation}\label{eq:defV_3}
    \begin{tikzpicture}
        \node[left] at (-2,0) {$-i\tilde{\mathbf{V}}_{(1,2)}(p,p')\equiv$};
        \draw[dashed] (-2,0)--(-0,0);
        \draw (0,0)--(1,1.732);
        \draw (0,0)--(1,-1.732);
        \draw[dashed] (0.625,-1.08625)--(0.625,1.08625);
        \draw[->] (-1,0)--(-0.905,0);
        \node[above] at (-1,0.05) {$q$};
        \draw[->] (0.835,1.44626)--(0.8375,1.45059);
        \draw[->] (0.8375,-1.45059)--(0.835,-1.44626);
        \draw[->] (0.31,0.5369)--(0.3125,0.54216);
        \draw[->] (0.3125,-0.54216)--(0.31,-0.5369);
        \draw[->] (0.625,-0.005)--(0.625,0);
        \node[left] at (0.836,1.45059) {$p'$};
        \node[left] at (0.31,0.54216) {$k'$};
        \node[left] at (0.836,-1.451) {$p$};
        \node[left] at (0.31,-0.54216) {$k$};
        \node[right] at (0.635,0) {$p-k$};
        \filldraw[black] (0,0) circle (0.05);
        \filldraw[black] (0.625,1.08625) circle (0.05);
        \filldraw[black] (0.625,-1.08625) circle (0.05);	
        \node at (2,-1.732) {$,$};
    \end{tikzpicture}
\end{equation}
where $p'\equiv p+q$ and $k'=k+q$. The correction to the vertex is then given by
\begin{equation}
    -i\mathbf{V}^{(1)}_{(1,2)}(p,p')=-ig_r-i\tilde{\mathbf{V}}_{(1,2)}(p,p')-i\delta_g.
\end{equation}
We now compute $-i\tilde{\mathbf{V}}_{(1,2)}(p,p')$ as defined in \eqref{eq:defV_3}:
\begin{equation}
    \begin{split}
        -i\tilde{\mathbf{V}}_{(1,2)}(p,p')=&-i(g_r\mu^\epsilon)^3\int\frac{d^4k}{(2\pi)^4}\frac{ie^{\frac{i\pi\epsilon}{2}}(\slashed{k}+m_r)}{(k^2-m_r^2+i\varepsilon)^{1+\frac{\epsilon}{2}}}\\
        &\times\frac{ie^{\frac{i\pi\epsilon}{2}}(\slashed{k}+\slashed{q}+m_r)}{((q+k)^2-m_r^2+i\varepsilon)^{1+\frac{\epsilon}{2}}}\\
        &\times\frac{ie^{\frac{i\pi\epsilon}{2}}}{((p-k)^2-M_r^2+i\varepsilon)^{1+\frac{\epsilon}{2}}}\\
        =&\,i(g_r\mu^\epsilon)^3\frac{\Gamma\left(3+\frac{3\epsilon}{2}\right)}{\Gamma\left(1+\frac{\epsilon}{2}\right)^3}\\
        &\times\int_0^1\int_0^{1-x}dxdyx^{\frac{\epsilon}{2}}y^{\frac{\epsilon}{2}}(1-x-y)^{\frac{\epsilon}{2}}\\
        &\times\int\frac{d^4l}{(2\pi)^4}\frac{ie^{\frac{i3\pi\epsilon}{2}}(l^2+N)}{(l^2-\Delta^2+i\varepsilon)^{3+\frac{3\epsilon}{2}}},
    \end{split}
\end{equation}
where 
\begin{subequations}
    \begin{align}
        l^{\mu}\equiv&\,k^\mu+xq^{\mu}-yp^\mu\\
        \begin{split}\Delta^2\equiv&\,m_r^2+(xq-yp)^2\\
					&-xq^2-yp^2+y(M_r^2-m_r^2)\end{split}\\
        N\equiv&\,[y\slashed{p}-x\slashed{q}+m_r][y\slashed{p}+(1-x)\slashed{q}+m_r].
    \end{align}
\end{subequations}
Then 
\begin{equation}
    \begin{split}
        -i\tilde{\mathbf{V}}_{(1,2)}(p,p')
        =&-\frac{i(g_r\mu^\epsilon)^3}{16\pi^2\Gamma(1+\frac{\epsilon}{2})^3}\\
        &\times\int_0^1\int_0^{1-x}dxdyx^{\frac{\epsilon}{2}}y^{\frac{\epsilon}{2}}(1-x-y)^{\frac{\epsilon}{2}}\\
        &\times\left(\frac{\Gamma\left(1+\frac{3\epsilon}{2}\right)N}{\left(\Delta^2-i\varepsilon\right)^{1+\frac{3\epsilon}{2}}}-\frac{2\Gamma\left(\frac{3\epsilon}{2}\right)}{\left(\Delta^2-i\varepsilon\right)^{\frac{3\epsilon}{2}}}\right).
    \end{split}
\end{equation}
Then, in the limit $\epsilon\ll1$, we have
\begin{equation}\label{eq:V_12infinieps}
    \begin{split}
        -i\tilde{\mathbf{V}}_{(1,2)}(p,p')=&-\frac{ig_r^3}{16\pi^2}\int_0^1\int_0^{1-x}dxdy\biggl(\frac{N}{\Delta^2-i\varepsilon}\\
        &+3-2\biggl(\log\left(\frac{\mu^2}{\Delta^2-i\varepsilon}\right)+\frac{2}{3\epsilon}-\gamma_E\biggr)\biggr)\\&+\mathcal{O}(\epsilon).
    \end{split}
\end{equation}
In order to cancel the $\epsilon^{-1}$ term and the Euler-Mascheroni constant, we set
\begin{equation}
    \delta_g=\frac{g_r^3}{8\pi^2}\left(\frac{2}{3\epsilon}-\gamma_E\right).
\end{equation}
Then, to NLO, we find the correction to the three point vertex to be
\begin{equation}\label{eq:finalresultV12yukawa}
    \begin{split}
        -i\mathbf{V}^{(1)}_{(1,2)}(p,p')=&-ig_r-\frac{ig_r^3}{16\pi^2}\int_0^1\int_0^{1-x}dydx\biggl(\frac{N}{\Delta^2-i\varepsilon}\\
        &+3-2\log\left(\frac{\mu^2}{\Delta^2-i\varepsilon}\right)\biggr).
    \end{split}
\end{equation}
Note that \eqref{eq:finalresultV12yukawa} has no poles in $\epsilon$, and is thus finite for all values of $\epsilon$.
\par Next we consider the diagrams with $N_\varphi=4$ and $N_\psi=0$, the correction to the four point vertex. Let us consider $\mathcal{V}_{1,2,3,4}$ defined as follows:
\begin{equation}\label{eq:defV4}
    \begin{tikzpicture}
        \node[left] at (-2.25,0) {$-i\mathcal{V}_{1,2,3,4}\equiv$};
        \draw (0,0) circle (1);
        \draw[dashed] (1.414,1.414)--(0.707,0.707);
        \draw[dashed] (-1.414,-1.414)--(-0.707,-0.707);
        \draw[dashed] (-1.414,1.414)--(-0.707,0.707);
        \draw[dashed] (1.414,-1.414)--(0.707,-0.707);
        \draw[->] (1.06419570569,1.06419570569)--(1.06066017178,1.06066017178);
        \draw[->] (-1.06419570569,-1.06419570569)--(-1.06066017178,-1.06066017178);
        \draw[->] (-1.06419570569,1.06419570569)--(-1.06066017178,1.06066017178);
        \draw[->] (1.06419570569,-1.06419570569)--(1.06066017178,-1.06066017178);
        \node[right] at (1.065,1.06066017178) {$p_4$};
        \node[right] at (1.065,-1.06066017178) {$p_1$};
        \node[left] at (-1.065,-1.0606017178) {$p_2$};
        \node[left] at (-1.065,1.06066017178) {$p_3$};
        \draw[->] (0.05,-1)--(0,-1);
        \draw[->] (-1,-0.05)--(-1,0);
        \draw[->] (-0.05,1)--(0,1);
        \draw[->] (1,0.05)--(1,0);
        \node[below,scale=0.9] at (0,-1) {$k+p_1$};
        \node[left,scale=0.9] at (-1.1,0) {$k+p_{1,2}$};
        \node[above,scale=0.9] at (0,1) {$k-p_4$};
        \node[right,scale=0.9] at (1,0) {$k$};
        \filldraw[black] (0.707,0.707) circle (0.05);
        \filldraw[black] (0.707,-0.707) circle (0.05);
        \filldraw[black] (-0.707,0.707) circle (0.05);
        \filldraw[black] (-0.707,-0.707) circle (0.05);
        \node[right] at (1.5,-1.414) {$\;,$};
    \end{tikzpicture}
\end{equation}
where $p_{1,2}\equiv p_1+p_2$. Let us notice that in considering a diagram of the form of $\mathcal{V}$, there are $4!$ different ways to contract the external scalars; however, for a given configuration of external momenta, due to the rotational symmetry of the diagram, $6$ of these diagrams are topologically equivalent. Then the NLO correction to the four-point vertex $-i\mathbf{V}_{(4,0)}(\{p_i\})$ is given by
\begin{equation}\label{eq:4pointoneloopscalarYukawa}
    \begin{split}
        -i\mathbf{V}^{(1)}_{(4,0)}(\{p_i\})=&-i\lambda_r-i\bigl(V(s)+V(t)+V(u)\bigr)\\
        &-i\bigl(\mathcal{V}_{1,2,3,4}+\mathcal{V}_{2,1,3,4}+\mathcal{V}_{1,3,2,4}\\
        &+\mathcal{V}_{4,2,3,1}+\mathcal{V}_{4,3,2,1}+\mathcal{V}_{3,2,1,4}\bigr)-i\delta_\lambda,
    \end{split}
\end{equation}
where $V(p^2)$ is defined in \eqref{eq:defV(p^2)}.  We now compute $-i\mathcal{V}_{1,2,3,4}$ as defined in \eqref{eq:defV4}:
\begin{equation}
    \begin{split}
        -i\mathcal{V}_{1,2,3,4}=&-\frac{g_r^4\mu^{4\epsilon}}{6}\int\frac{d^4k}{(2\pi)^4}\tr\biggl(\frac{ie^{\frac{i\pi\epsilon}{2}}(\slashed{k}+m_r)}{(k^2-m_r^2)^{1+\frac{\epsilon}{2}}}\\
        &\times\frac{ie^{\frac{i\pi\epsilon}{2}}(\slashed{k}+\slashed{p}_1+m_r)}{((k+p_1)^2-m_r^2)^{1+\frac{\epsilon}{2}}}\\
        &\times\frac{ie^{\frac{i\pi\epsilon}{2}}(\slashed{k}+\slashed{p}_{1,2}+m_r)}{((k+p_{1,2})^2-m_r^2)^{1+\frac{\epsilon}{2}}}\\
        &\times\frac{ie^{\frac{i\pi\epsilon}{2}}(\slashed{k}-\slashed{p}_4+m_r)}{((k-p_1)^2-m_r^2)^{1+\frac{\epsilon}{2}}}\biggr)\\
        =&\,\frac{2ig_r^4\mu^{4\epsilon}e^{2\pi i\epsilon}}{3}\int_0^1\int_0^{1-x}\int_0^{1-x-y}dxdydz\\
        &\times\frac{\Gamma\left(4+2\epsilon
        \right)}{\Gamma\left(1+\frac{\epsilon}{2}\right)^4}x^{\frac{\epsilon}{2}}y^{\frac{\epsilon}{2}}z^{\frac{\epsilon}{2}}(1-x-y-z)^{\frac{\epsilon}{2}}\\
        &\times\int\frac{d^4l}{(2\pi)^4}\frac{i\left((l^2)^2+N_{1,2,3,4}l^2+\tilde{N}_{1,2,3,4}\right)}{\left(l^2-\Delta^2_{1,2,3,4}+i\varepsilon\right)^{4+2\epsilon}},
    \end{split}
\end{equation}
where
\begin{equation}
    \begin{split}
        l^\mu\equiv k^\mu+xp_1^\mu+yp_{1,2}^\mu-zp_4^\mu
    \end{split}
\end{equation}
and
\begin{equation}
    \begin{split}
        \Delta^2_{1,2,3,4}\equiv&\,m_r^2+(xp_1+yp_{1,2}-zp_4)^2\\&-(xp_1^2+yp_{1,2}^2+zp_4^2).
    \end{split}
\end{equation}
Let us define
\begin{subequations}
    \begin{align}
        q_1\equiv&\,zp_4-xp_1-yp_{1,2}\\
        q_2\equiv&\,q_1+p_1\\
        q_3\equiv&\,q_1+p_{1,2}\\
        q_4\equiv&\,q_1-p_4.
    \end{align}
\end{subequations}
Then, we define
\begin{equation}
    \begin{split}
        N_{1,2,3,4}\equiv&\,6m_r^2+q_1\cdot q_2+q_3\cdot q_4+q_2\cdot q_3+q_1\cdot q_4\\&-\frac{1}{2}(q_1\cdot q_3+q_2\cdot q_4)
    \end{split}
\end{equation}
and 
\begin{equation}
    \begin{split}
        \tilde{N}_{1,2,3,4}\equiv&\,m_r^4+m_r^2(q_1\cdot q_2+q_1\cdot q_3+q_1\cdot q_4\\
        &+q_2\cdot q_3+q_2\cdot q_4+q_3\cdot q_4\bigr)\\
        &+(q_1\cdot q_2)(q_3\cdot q_4)-(q_1\cdot q_3)(q_2\cdot q_4)\\&+(q_1\cdot q_4)(q_2\cdot q_3).
    \end{split}
\end{equation}
Then, we have 
\begin{equation}
    \begin{split}
        -i\mathcal{V}_{1,2,3,4}=&\,\frac{ig_r^4\mu^{4\epsilon}}{24\pi^2\Gamma\left(1+\frac{\epsilon}{2}\right)^4}\int_0^1\int_0^{1-x}\int_0^{1-x-y}dxdydz\\
        &\times x^{\frac{\epsilon}{2}}y^{\frac{\epsilon}{2}}z^{\frac{\epsilon}{2}}(1-x-y-z)^{\frac{\epsilon}{2}}\\
        &\times(\Delta^2_{1,2,3,4}-i\varepsilon)^{-2\epsilon}\biggl(6\Gamma(2\epsilon)\\
        &-\frac{2N_{1,2,3,4}\Gamma(1+2\epsilon)}{\Delta^2_{1,2,3,4}-i\varepsilon}+\frac{\tilde{N}_{1,2,3,4}\Gamma(2+2\epsilon)}{(\Delta^2_{1,2,3,4}-i\varepsilon)^2}\biggr).
    \end{split}
\end{equation}
Hence, taking the $\epsilon\ll1$ limit, we find 
\begin{equation}\label{eq:V1234infeps}
    \begin{split}
        -i\mathcal{V}_{1,2,3,4}=&\,\frac{ig_r^4}{24\pi^2}\int_0^1\int_0^{1-x}\int_0^{1-x-y}dxdydz\\
        &\times\biggl(\frac{\tilde{N}_{1,2,3,4}}{(\Delta^2_{1,2,3,4}-i\varepsilon)^2}-\frac{2N_{1,2,3,4}}{\Delta^2_{1,2,3,4}-i\varepsilon}\\
        &+6\biggl[\frac{1}{2\epsilon}-\gamma_E+\log\left(\frac{\mu^2}{\Delta^2_{1,2,3,4}-i\varepsilon}\right)\biggr]\\
        &-\frac{11}{6}\biggr)+\mathcal{O}(\epsilon).
    \end{split}
\end{equation}
In order to cancel the $\epsilon^{-1}$ term and Euler-Mascheroni constant from \eqref{eq:V(p^2)fracreg.infiniteep} and \eqref{eq:V1234infeps} in \eqref{eq:4pointoneloopscalarYukawa}; we set the shift $\delta_\lambda$ to be
\begin{equation}
    \begin{split}
        \delta_\lambda=&-\frac{3\lambda_r^2}{32\pi^2}\biggl(\frac{1}{\epsilon}-\gamma_E\biggr)+\frac{3g_r^4}{2\pi^2}\left(\frac{1}{2\epsilon}-\gamma_E\right).
    \end{split}
\end{equation}
Define 
\begin{equation}
    \begin{split}
        \tilde{D}_{ijkl}\equiv&\int_0^1\int_0^{1-x}\int_0^{1-x-y}dxdydz\\
        &\times\biggl(\frac{\tilde{N}_{i,j,k,l}}{(\Delta^2_{i,j,k,l}-i\varepsilon)^2}-\frac{2N_{i,j,k,l}}{\Delta^2_{i,j,k,l}-i\varepsilon}\\
        &+6\log\left(\frac{\mu^2}{\Delta^2_{i,j,k,l}-i\varepsilon}\right)+\frac{11}{6}\biggr).
    \end{split}
\end{equation}
Then, the NLO correction to $\mathbf{V}_{(4,0)}(\{p_i\})$ is:
\begin{equation}\label{eq:Finalresultfourpointscalaryukawa}
    \begin{split}
        &-i\mathbf{V}^{(1)}_{(4,0)}(\{p_i\})=-i\lambda_r\\
        &-\frac{i\lambda_r^2}{32\pi^2}\int_0^1dx\biggl[\log\left(\frac{\mu^2}{M_r^2-x(x-1)s-i\varepsilon}\right)-6\\
        &+\log\left(\frac{\mu^2}{M_r^2-x(x-1)t}\right)+\log\left(\frac{\mu^2}{M_r^2-x(x-1)u}\right)\biggr]\\
        &+\frac{ig_r^4}{24\pi^2}\sum_{\sigma\in\mathcal{A}_{4}}\tilde{D}_{\sigma(1234)}\,,
    \end{split}
\end{equation}
where $\mathcal{A}_{4}$ is the set of all non-cyclic permutations of $\{1,2,3,4\}$\footnote{Often referred to as the alternating group over four letters.}. \eqref{eq:Finalresultfourpointscalaryukawa} has no poles in $\epsilon$, and thus is finite for all values of $\epsilon$. 
\par Through computing the superficial degree of divergence of Yukawa theory, we identified all the superficially divergent diagrams of the theory. We then showed, to NLO, that SAR self-consistently regularizes Yukawa theory, yielding finite results for all superficially divergent diagrams at NLO. We can compare \eqref{eq:YukawaScalarSelfEnergy}, \eqref{eq:Yukawafermionselfenergy}, \eqref{eq:finalresultV12yukawa}, and \eqref{eq:Finalresultfourpointscalaryukawa} to standard textbook results (e.g. \cite{Peskin:1995ev,Srednicki:2007qs}\footnote{Note that \cite{Srednicki:2007qs} considers pseudo-scalar Yukawa theory.}) and see that they agree up to renormalization scheme dependent constant terms such as $\gamma_E$ and $\log4\pi$. 
\pagebreak
\section{Conclusions and Outlook}
In this work, we introduced a novel, mathematically rigorous regularization scheme for quantum field theories, which we term systematic analytic regularization (SAR). Building upon the analytic techniques developed by Riesz and Hadamard, this approach systematically modifies the kinetic terms in the action by analytically continuing the power of the kinetic operators from $1$ to $1 + \epsilon/2$. As a result, the divergences in loop integrals manifest as poles in the regulator $\epsilon$, similarly to dimensional regularization, but within a framework that maintains fixed spacetime dimensionality. 
\par Crucially, systematic analytic regularization can be understood as a consistent, action-level generalization of analytic regularization, leading to a new class of Feynman rules. In regularizing at the level of the action, we avoid manipulations of any formally divergent quantities while ensuring that all symmetries of the theory are preserved. We applied this scheme to two representative quantum field theories: $\varphi^4$ theory and Yukawa theory.
\par Our study began by examining the symmetries of the unregularized theories before applying SAR to compare their behavior. We then analyzed the superficial degrees of divergence to identify divergent diagrams.
\par A central result was the demonstration that SAR preserves all symmetries (at the classical level) and makes all results finite at NLO. It can be shown (see e.g. \cite{Peskin:1995ev}) that dim reg also renders the theories that we studied finite at NLO. However, SAR has a critical advantage: it avoids the ambiguities inherent in dimensional regularization that arise from mismatches between the number of spacetime dimensions and the representations of fields (e.g., the ill-defined nature of $\gamma^5$). By keeping the spacetime dimension fixed, SAR maintains clarity in Clifford algebra and trace identities, especially relevant in theories sensitive to chiral and axial structures.
\par Much like dim reg, SAR also introduces a fictitious scale $\mu$, indicating that the Callan-Symanzik equations apply to SAR. Despite their conceptual differences, the practical implementation of SAR is no more difficult than dim reg, and the integrals produced are of comparable complexity. In addition, in dim reg, singularities always arise from a Gamma function of the form $\Gamma(-n + \epsilon/2)$, whereas in SAR, the pole structure depends on the number of propagators in the loop: $\Gamma(-n + P\epsilon/2)$, where $P$ is the number of propagators. This distinction could have deeper implications for understanding the structure of divergences in multi-loop processes. Overall, SAR offers a promising alternative to dimensional regularization, with clearer conceptual foundations, particularly in contexts where preserving spacetime dimension and internal symmetries is essential.
\par Looking forward, a natural and significant next step is to extend this method to QED, with the goal of constructing a version of SAR that preserves BRST invariance and SUSY both classically and quantum mechanically, while also accurately reproducing the axial anomaly. If successful, this would establish SAR as a powerful tool for gauge theories, paving the way for applications to non-Abelian Yang-Mills theories. The ultimate goal is to show that SAR is a regularization scheme that simultaneously preserves Lorentz invariance, gauge invariance, and supersymmetry.
\par Finally, although our analysis was performed entirely within the path integral formalism, it would be of great theoretical interest to explore how SAR can be formulated and interpreted within the canonical (Hamiltonian) framework. Such an investigation could yield further insight into the non-local structures introduced by the method and deepen our understanding of its implications in both perturbative and non-perturbative settings.

\section{Acknowledgments}
We would like to thank the NRF and SA-CERN for the generous support and 
Robert D. Pisarski for the original idea to analytically continue the power of the kinetic operator in the Lagrangian. We would also like to thank Saswato Sen for providing us with valuable resources regarding analytically continued kinetic operators in the context of condensed matter physics. 
\par This research was conducted in part by WAH while visiting the Okinawa Institute of
Science and Technology (OIST) through the Theoretical Sciences Visiting Program (TSVP).
\bibliography{biblio}
\bibliographystyle{iopart-num}
\end{document}